\documentclass[lettersize,journal]{IEEEtran}
\usepackage{amsmath,amssymb,amsfonts}
\usepackage{algorithmic}
\usepackage{algorithm}
\usepackage{array}
\usepackage{graphicx}
\usepackage[caption=false,font=footnotesize,labelfont=rm,textfont=rm]{subfig}
\usepackage{textcomp}
\usepackage{stfloats}
\usepackage{booktabs}
\usepackage{subfig}
\usepackage{subfloat}
\usepackage{multirow}
\usepackage{multicol}
\usepackage{stfloats}
\usepackage{siunitx}
\usepackage{url}
\usepackage{verbatim}
\usepackage{graphicx}
\usepackage{amsmath}
\usepackage{bbding}
\usepackage[algo2e]{algorithm2e}
\usepackage{cite}
\usepackage{tcolorbox}
\usepackage{bbding}
\usepackage{hyperref}

\usepackage{tcolorbox}
\tcbuselibrary{skins,breakable}
\usepackage{listings}
\usepackage{xcolor}
\usepackage{enumitem}

\usepackage{subcaption}
\usepackage{pifont} 

\usepackage{makecell}
\usepackage{booktabs}
\usepackage{colortbl}

\usepackage{pifont}
\usepackage{color}

\lstdefinestyle{promptjson}{
    basicstyle=\ttfamily\scriptsize,
    breaklines=true,
    breakatwhitespace=false,
    columns=fullflexible,
    keepspaces=true,
    showstringspaces=false,
    frame=none,
    backgroundcolor=\color{gray!5},
    xleftmargin=0.5em,
    xrightmargin=0.5em
}

\usepackage{xcolor}
\usepackage{enumitem}
\usepackage{tcolorbox}
\tcbuselibrary{breakable}

\newtcolorbox{promptbox}[1]{
    breakable,
    colback=white,
    colframe=gray!75!black,
    colbacktitle=gray!75!black,
    coltitle=white,
    title={#1},
    fonttitle=\large\bfseries,
    boxrule=0.8pt,
    arc=2mm,
    left=8pt,
    right=8pt,
    top=6pt,
    bottom=6pt,
    toptitle=6pt,
    bottomtitle=6pt,
    width=\textwidth
}

\usepackage{threeparttable} 
\begin{document}

\title{DepressionAgent: Reading, Listening, Seeing, and Deliberating Multimodal Evidence for Depression Risk Assessment}

\author{Fangjie Zhu, Haifeng Lu, Sicheng Zhao, Runhao Zeng, and Xiping Hu
\thanks{Fangjie Zhu, Runhao Zeng, and Xiping Hu are with Guangdong-Hong Kong-Macao Joint Laboratory for Emotional Intelligence and Pervasive Computing, Shenzhen MSU-BIT University, Shenzhen 518107, China (e-mail: e1351405@u.nus.edu, runhaozeng.cs@gmail.com,  huxp@smbu.edu.cn).}
\thanks{Haifeng Lu is with the Artificial Intelligence Research Institute, Shenzhen MSU-BIT University, Shenzhen 518107, China, and the Department of Electrical and Computer Engineering, The University of Hong Kong, China (e-mail: luhfhku@hku.hk)}
\thanks{Sicheng Zhao is with the Department of Psychological and Cognitive Sciences, Tsinghua University, Beijing. (e-mail: schzhao@tsinghua.edu.cn).}


\thanks{Fangjie Zhu and Haifeng Lu contributed equally to this work.}
\thanks{Sicheng Zhao and Runhao Zeng are corresponding authors.}
}

\markboth{}%
{Shell \MakeLowercase{\textit{et al.}}: A Sample Article Using IEEEtran.cls for IEEE Journals}


\maketitle

\begin{abstract}
Multimodal depression risk assessment requires jointly interpreting textual, acoustic, and visual cues that are often subtle, non-specific, context-dependent, and potentially inconsistent across modalities. Existing multimodal approaches predominantly learn latent representations through feature fusion, leaving the evidence underlying a prediction and the treatment of cross-modal disagreement largely implicit. We propose \textbf{DepressionAgent}, an evidence-centric agentic framework that transforms multimodal depression assessment from implicit feature fusion into explicit evidence deliberation. DepressionAgent first converts textual, acoustic, and visual inputs into modality-specific evidence, and then organizes self-report and behavioral evidence into parallel support--challenge deliberation branches. Cross-modal arbitration explicitly examines agreement and disagreement between the two branches, with conflict reflection revisiting inconsistent assessments before decision making. A subsequent risk reflection mechanism provides an independent textual second opinion for initially low-risk cases to reduce potentially missed risk signals. Without depression-specific supervised training or parameter fine-tuning, DepressionAgent achieves competitive performance on multiple public benchmarks. Extensive ablations, cross-model evaluations, qualitative analyses, and clinician assessments further demonstrate the effectiveness and inspectability of the proposed framework.
\end{abstract}

\begin{IEEEkeywords}
Multimodal Depression Risk Assessment, Affective Computing, Multi-agent Systems, Evidence Reasoning
\end{IEEEkeywords}

\section{Introduction}
\label{sec:introduction}

\begin{figure}[t]
\centering
\includegraphics[width=\linewidth]{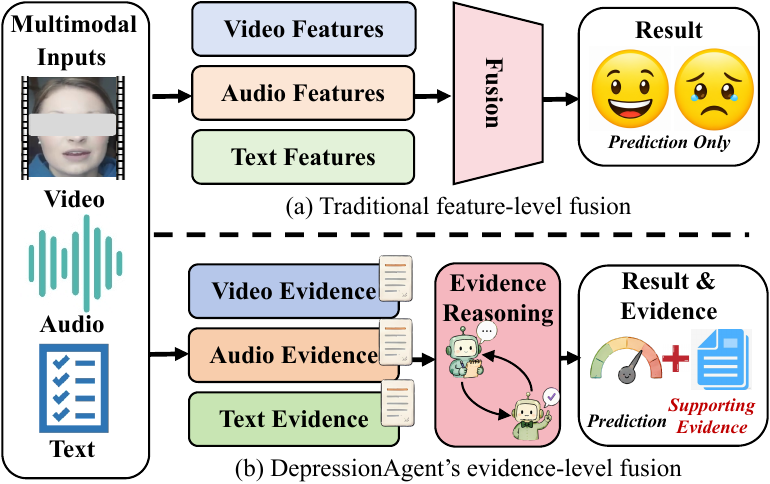}
\caption{
Conceptual comparison between conventional feature-level fusion and the evidence-centric reasoning paradigm of DepressionAgent. Conventional approaches integrate multimodal information in latent representations before prediction, whereas DepressionAgent explicitly organizes modality-specific evidence and reasons over their relationships to jointly produce the risk assessment and its supporting evidence.
}
\label{fig:paradigm}
\end{figure}

\IEEEPARstart{D}{epression} has become a major public health concern, substantially impairing individuals' quality of life and imposing considerable social and economic burdens. With recent advances in digital mental health and multimodal artificial intelligence, automatic depression risk assessment from textual, acoustic, and visual data collected in natural interaction scenarios, such as clinical interviews and video blogs, has emerged as an important research topic in mental health screening and multimodal affective computing \cite{chen2026towards, wang2025ai, he2022deep}. Text conveys subjective experiences and affective content, speech reveals expressive characteristics such as prosody, speaking rate, and pauses, and video captures nonverbal behaviors including facial expressions, gaze, and body movement. Multimodal analysis therefore offers the potential to characterize depression-related affective and behavioral states from complementary perspectives \cite{alghowinem2016multimodal}.

Multimodal depression risk assessment, however, is not a conventional pattern recognition problem determined by a small number of salient features. Depression-related cues are often subtle, non-specific, and highly dependent on context and individual differences \cite{cummins2015review}. Negative language, low-energy speech, or reduced facial expressiveness may be associated with depression risk, but may also arise from transient stress, fatigue, personal communication style, or recording conditions. Moreover, different modalities may provide inconsistent observations. An individual may repeatedly describe negative experiences while exhibiting relatively natural vocal and visual behaviors; conversely, noticeable nonverbal changes may occur without explicit negative self-reports. The central challenge is therefore not simply to extract more unimodal features, but to determine how distributed and potentially conflicting information should be interpreted together \cite{hussain2026depression, charlin2007scripts, croskerry2009universal}.

Most existing multimodal depression assessment methods address this problem through increasingly powerful modality encoders and fusion architectures. Textual, acoustic, and visual signals are encoded into latent representations and subsequently integrated through attention mechanisms, Transformers, graph models, or other fusion networks\cite{yang2017multimodal}. Although effective for prediction, this paradigm compresses multimodal observations and their interactions into high-dimensional representations. Consequently, which observations support or oppose a prediction, whether apparently relevant cues admit alternative explanations, and how disagreement between modalities affects the final decision are generally left implicit. As illustrated in Fig.~\ref{fig:paradigm}, the limitation is therefore not simply insufficient multimodal interaction, but that the assessment is performed primarily through \emph{feature-level fusion rather than explicit evidence-level reasoning}.

We argue that multimodal depression assessment can instead be viewed as a process of \emph{evidence deliberation}. Modality-specific observations should first be preserved as explicit evidence rather than prematurely fused. Potentially risk-related evidence should be examined together with counter-evidence and contextual alternatives because many depression-related manifestations are non-specific. Moreover, disagreement between self-reported experiences and observable behaviors should itself become an explicit object of reasoning rather than being absorbed into a joint representation. Finally, because subtle self-reported symptoms may remain insufficiently reflected in an initially low-risk multimodal decision, such decisions may benefit from an additional risk-oriented review.

Following this perspective, we propose \textbf{DepressionAgent}, an \textbf{evidence-centric and reflection-driven agentic framework} for multimodal depression risk assessment. DepressionAgent first transforms textual, acoustic, and visual inputs into modality-specific evidence while separating observation from risk interpretation. The resulting evidence is organized into a self-report branch and an audio-visual behavioral branch. Within each branch, a Hypothesis Agent performs risk discovery, while a Challenger Agent actively searches for counter-evidence and plausible alternative explanations before a branch-level decision is formed. The two branch assessments are subsequently reconciled through cross-modal arbitration. When disagreement is detected, \emph{Conflict Reflection} identifies the source of inconsistency and returns it to the deliberation process for reconsideration. After the initial multimodal decision, \emph{Risk Reflection} provides an independent textual second opinion for every initially low-risk case, targeting potentially overlooked self-reported risk evidence.

The resulting framework is not characterized simply by the number of agents, but by how evidence is represented, challenged, exchanged, and revised. Modality-isolated observation preserves heterogeneous information sources; support--challenge deliberation explicitly addresses the low specificity of depression-related cues; Conflict Reflection turns cross-modal disagreement into a revisable reasoning state; and Risk Reflection provides a dedicated safeguard against residual false-negative decisions. Together, these mechanisms establish an explicit reasoning path from multimodal observations to evidence deliberation and final assessment.

DepressionAgent requires no depression-specific supervised training or parameter fine-tuning. We evaluate the framework on two public in-the-wild multimodal depression benchmarks. Functional and sub-agent ablations examine the contribution of the proposed reasoning operations, while prediction-transition analysis directly quantifies how Conflict Reflection and Risk Reflection revise intermediate decisions. Cross-foundation-model experiments evaluate the transferability of the framework. In addition, explicit-language masking, qualitative case analysis, and clinician evaluation examine whether the resulting evidence and reasoning are informative and inspectable beyond the final prediction.

The main contributions of this work are summarized as follows:

\begin{enumerate}
    \item We introduce an \textbf{evidence-centric formulation} of multimodal depression risk assessment that shifts the focus from latent feature fusion to explicit deliberation over supporting evidence, counter-evidence, contextual alternatives, and cross-modal disagreement.

    \item We propose \textbf{DepressionAgent}, a reflection-driven agentic framework that combines modality-isolated observation, support--challenge branch deliberation, cross-modal arbitration with Conflict Reflection, and low-risk-oriented Risk Reflection to progressively examine and revise multimodal evidence.

    \item DepressionAgent requires no depression-specific supervised training or parameter fine-tuning. Extensive experiments, including matched-protocol benchmark comparisons, functional and sub-agent ablations, reflection analysis, cross-foundation-model evaluation, language masking, qualitative analysis, and clinician assessment, demonstrate its predictive capability and provide evidence for the effectiveness and inspectability of the proposed reasoning process.
\end{enumerate}

\section{Related Work}

\subsection{Multimodal Depression Assessment}

Multimodal depression assessment has long been motivated by the premise that behavioral signals from multiple sources can improve risk recognition. Classical studies have established representation, alignment, and fusion as central problems in multimodal machine learning\cite{baltrusaitis2018multimodal,baltrusaitis2019survey,10508050,Shangguan2025FacialSurvey}. Progress in depression assessment has been supported by controlled clinical-interview datasets such as DAIC-WOZ and AVEC\cite{gratch2014daic,ringeval2019avec}, as well as in-the-wild benchmarks such as D-Vlog\cite{yoon2022dvlog}. Recent work has also revealed potential dataset-specific biases, including shortcuts associated with interviewer prompts in DAIC-WOZ\cite{burdisso2024daic}.

Methodologically, approaches have evolved from early and late fusion toward Transformer-based cross-modal modeling, spatiotemporal attention, and graph-based architectures\cite{tao2024depmstat,shen2024multimodal}. Despite these advances, most methods follow a common paradigm: modality-specific inputs are encoded into latent representations, interactions are learned through fusion, and the resulting representation is mapped to a depression prediction. Thus, the supporting, contradictory, and alternative evidence underlying a decision is generally not treated as an explicit reasoning object. DepressionAgent instead treats textual, acoustic, and visual observations as explicit evidence and focuses on how heterogeneous evidence can be interpreted, challenged, reconciled, and revised.

\subsection{LLM-Based and Explainable Depression Assessment}

Large language models have increasingly been explored for medical reasoning and decision support, while the broader explainable AI literature has emphasized the distinction between post-hoc explanations and inherently interpretable reasoning\cite{singhal2023medpalm,rudin2019stop,longo2024explainable}. These developments have recently extended to depression assessment, where LLMs are used to extract clinically meaningful indicators and generate interpretable intermediate representations. Early efforts used LLMs to extract depression-related indicators from transcripts and combine them with facial features for multimodal prediction\cite{sadeghi2024harnessing}, while recent multimodal LLMs and summary-based approaches further integrate audio-visual information or use LLM-generated clinical summaries to guide multimodal fusion\cite{zhao2025ithearss,teng2026dynamic}.

Recent work has explored explainable multimodal depression recognition through structured symptom summaries and underlying-cause generation\cite{zheng2025emdr}. Related approaches progressively enrich LLM-generated clinical summaries to guide multimodal fusion and produce interpretable reports\cite{teng2026dynamic}. These approaches substantially improve the visibility of clinically meaningful information; however, their explanations are primarily formulated as symptom summarization or summary-guided fusion rather than explicit evidence deliberation.

Dep-LLM\cite{lyu2026depllm} further demonstrates the feasibility of training-free, evidence-guided depression reasoning by decomposing clinical interview transcripts into clinically aligned factors, generating evidence-grounded rationales, and incorporating uncertainty into the final aggregation. This work establishes structured textual reasoning as a promising direction, but remains centered on transcript-based evidence. In contrast, DepressionAgent addresses heterogeneous evidence natively observed from text, audio, and video, where modality-specific observations may be non-specific or contradictory. Rather than only weighting or summarizing evidence, it explicitly performs support--challenge deliberation, cross-modal arbitration, and feedback-based reflection.

\subsection{Agentic Reasoning for Mental Health Assessment}

Agentic LLMs have also been introduced into psychiatric assessment, moving beyond direct prediction toward role-specialized reasoning and structured clinical workflows. MAGI~\cite{bi2025magi} transforms the MINI psychiatric interview into a multi-agent process involving interview navigation, adaptive questioning, judgment, and diagnosis, demonstrating the potential of coordinated agents for structured psychiatric assessment. More recent systems such as WiseMind~\cite{wu2026wisemind} further employ knowledge-guided multi-agent collaboration for psychiatric diagnosis. These studies establish that structured psychiatric agents and multi-agent clinical reasoning are increasingly viable, but primarily focus on active interviewing, diagnostic criteria, knowledge-guided reasoning, or textual clinical information.

DepressionAgent instead focuses on the \emph{evidence deliberation process} for naturally occurring multimodal observations. Textual self-report, acoustic expression, and visual behavior are independently examined and challenged before cross-modal arbitration. When evidence conflicts, the disagreement itself becomes a reasoning object and can trigger reflection and revision. Thus, our framework differs from prior agentic systems not primarily by introducing multiple specialized agents, but by organizing their collaboration around the independent examination, reconciliation, and reflection of heterogeneous multimodal evidence.

\section{Method}
\label{sec:method}

\subsection{Problem Formulation}

Given an input video $X$, we extract its transcript $T$, audio signal $A$, and visual stream $V$ as multimodal observations, and aim at predicting the dataset-defined binary depression risk label $Y$.

Conventional multimodal depression risk assessment methods typically learn a direct mapping from multimodal observations to the final prediction using modality-specific encoders and feature-fusion networks:
\begin{equation}
    \hat{Y}=F_{\theta}(T,A,V),
\end{equation}
where $F_{\theta}$ denotes a multimodal classification model parameterized by $\theta$, and $\hat{Y}$ is the predicted label. In this paradigm, modality-specific information and cross-modal interactions are encoded in latent representations, making the evidence underlying a prediction and the treatment of cross-modal disagreement difficult to inspect explicitly.

We instead formulate multimodal depression risk assessment as an \textbf{evidence-deliberation process}:
\begin{equation}
    \left(\hat{Y},\mathcal{C}\right)
    =
    \mathcal{R}_{\mathrm{evid}}(T,A,V),
\end{equation}
where $\mathcal{R}_{\mathrm{evid}}$ denotes the proposed evidence-centric agentic reasoning process and $\mathcal{C}$ denotes the structured reasoning record generated during assessment. Rather than directly mapping multimodal observations to a prediction, DepressionAgent explicitly represents modality-specific evidence and progressively performs evidence discovery, counter-analysis, cross-modal arbitration, and reflection.

\subsection{Evidence-Centric Agentic Reasoning}

\begin{figure*}[t]
\centering
\includegraphics[width=\linewidth]{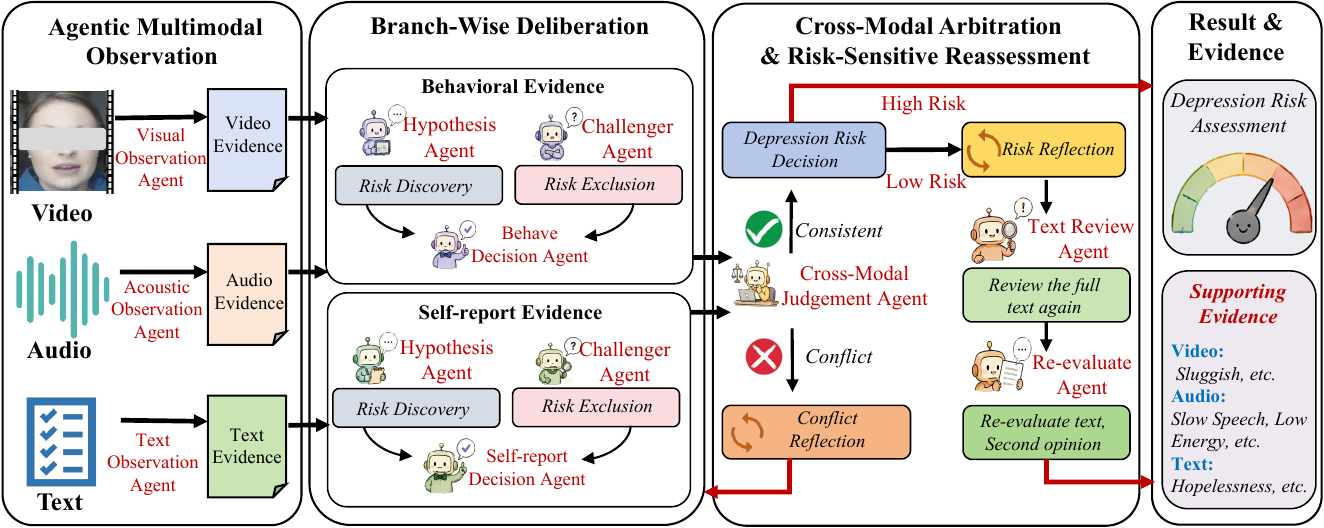}
\caption{
Overview of DepressionAgent. Multimodal observations are first transformed into modality-specific evidence and independently deliberated within self-report and behavioral branches. Cross-modal arbitration examines the relationship between the resulting assessments, with disagreement activating Conflict Reflection. Initially low-risk decisions subsequently activate Risk Reflection for an independent textual second opinion. The final output contains both the depression risk assessment and its supporting evidence.
}
\label{fig:framework}
\end{figure*}

We propose \textbf{DepressionAgent}, an evidence-centric and reflection-driven agentic framework for multimodal depression risk assessment. Its central design principle is to make explicit evidence, rather than latent multimodal features or isolated agent predictions, the basic object of reasoning. As illustrated in Fig.~\ref{fig:framework}, DepressionAgent progressively transforms multimodal observations into structured evidence, deliberates over competing interpretations, examines disagreement across modalities, and revises intermediate decisions through reflection.

The framework consists of four stages. \textbf{Agentic Multimodal Observation} converts textual, acoustic, and visual inputs into modality-specific evidence while separating observation from risk interpretation. \textbf{Branch-Wise Deliberation} organizes the evidence into a self-report branch and an audio-visual behavioral branch, where Hypothesis and Challenger Agents perform complementary risk discovery and risk exclusion before branch-level decisions are formed. \textbf{Cross-Modal Evidence Arbitration} compares the resulting branch assessments and explicitly examines whether their underlying evidence is compatible or conflicting; detected disagreement activates Conflict Reflection and feeds the identified issue back into deliberation. Finally, \textbf{Risk-Sensitive Textual Reassessment} re-examines every initially low-risk case through an independent textual second-opinion pathway.

These stages introduce two complementary reflection mechanisms. Conflict Reflection is disagreement-driven: when self-report and behavioral evidence support different assessments, the framework explicitly identifies the source of inconsistency and revisits the corresponding reasoning. Risk Reflection is decision-driven: an initially low-risk multimodal prediction triggers an independent review of the complete transcript to reduce potentially missed self-reported risk signals. Together, they move multimodal depression assessment beyond one-pass aggregation toward iterative evidence deliberation.

\subsection{Agentic Multimodal Observation}

Depression-related cues are heterogeneous across modalities, and premature cross-modal interaction may cause information from one modality to influence the interpretation of another. We therefore separate modality-specific observation from subsequent risk reasoning. At this stage, each observation agent accesses only its assigned modality and is explicitly instructed to describe observable information without predicting the depression risk label. This preserves modality-specific information boundaries and provides independent evidence for downstream deliberation.

Specifically, DepressionAgent assigns an independent observation agent to each modality:
\begin{equation}
(E_T,E_A,E_V)
=
\left(
\mathcal{A}_{T}^{\mathrm{obs}}(T),
\mathcal{A}_{A}^{\mathrm{obs}}(A),
\mathcal{A}_{V}^{\mathrm{obs}}(V)
\right),
\end{equation}
where $\mathcal{A}_{T}^{\mathrm{obs}}$, $\mathcal{A}_{A}^{\mathrm{obs}}$, and $\mathcal{A}_{V}^{\mathrm{obs}}$ denote the textual, acoustic, and visual observation agents, respectively, and $E_T$, $E_A$, and $E_V$ denote the corresponding modality-specific evidence.

To maintain consistent evidence extraction while preserving modality-specific reasoning boundaries, each observation agent follows a predefined observation scope, as summarized in Table~\ref{tab:observation_scope}.

\begin{table}[t]
    \centering
    \caption{Modality-specific observation scopes of the observation agents.}
    \label{tab:observation_scope}
    \setlength{\tabcolsep}{1mm}
    \renewcommand{\arraystretch}{1.15}
    \fontsize{9}{9}\selectfont
    \begin{tabular}{@{}p{0.17\columnwidth}p{0.76\columnwidth}@{}}
        \toprule
        \textbf{Agent} & \textbf{Observation Scope} \\
        \midrule

        Visual &
        Facial expressivity; gaze engagement; psychomotor activity;
        posture dynamics; response timing. \\

        Acoustic &
        Prosodic variability; speech timing dynamics; pause and hesitation
        patterns; vocal energy dynamics; adaptation patterns. \\

        Text &
        Internal states; behavioral patterns; cognitive framing;
        protective factors; contextual qualifiers. \\

        \bottomrule
    \end{tabular}
\end{table}

\textbf{Visual Observation Agent.}
The visual observation agent examines nonverbal behavior throughout the video and converts observable behavioral patterns into structured visual evidence. It focuses on what is visually present while avoiding depression-related interpretation of individual behavioral cues.

\textbf{Acoustic Observation Agent.}
The acoustic observation agent analyzes how the subject speaks and summarizes observable paralinguistic patterns as acoustic evidence. Its role is to characterize speech behavior rather than infer depression risk from isolated acoustic characteristics.

\textbf{Text Observation Agent.}
The textual observation agent examines the complete transcript and summarizes self-reported information relevant to subsequent risk reasoning. Unlike behavioral modalities, text directly conveys semantically explicit internal experiences and contextual statements. Representative expressions are therefore retained when necessary so that downstream agents can inspect the original self-report evidence.

\subsection{Branch-Wise Deliberation}

Depression-related cues are often subtle and non-specific. Similar observations may arise from fatigue, temporary stress, individual expressive style, or recording conditions. A potentially relevant cue should therefore not be directly converted into a risk judgment without considering counter-evidence and plausible alternative explanations. DepressionAgent addresses this characteristic through a support--challenge--decision protocol rather than treating the first plausible interpretation as the branch conclusion.

Textual and nonverbal observations also play different semantic roles in depression assessment. Text primarily conveys self-reported experiences and contextual information, whereas audio and video jointly characterize how an individual speaks and behaves through observable vocal and visual signals. We therefore maintain an independent textual branch while combining acoustic and visual evidence into an audio-visual behavioral branch:
\begin{equation}
    E_{\mathrm{AV}}=\left(E_A,E_V\right).
\end{equation}

For each branch $b\in\{T,\mathrm{AV}\}$, DepressionAgent applies the following deliberative process.

\textbf{Hypothesis Agent.}
Given branch evidence $E_b$, the Hypothesis Agent actively identifies observations that may support depression risk and constructs an initial interpretation:
\begin{equation}
H_b =
\mathcal{A}_{b}^{\mathrm{hyp}}(E_b).
\end{equation}
Its purpose is to comprehensively surface potentially relevant risk evidence rather than immediately produce a branch-level decision.

\textbf{Challenger Agent.}
The Challenger Agent critically examines the initial hypothesis and actively searches for information that weakens or contradicts it, including protective evidence, contextual factors, and plausible non-depressive explanations:
\begin{equation}
Q_b =
\mathcal{A}_{b}^{\mathrm{cha}}(E_b,H_b).
\end{equation}
This counter-analysis is particularly important because many depression-related cues are non-specific and may admit reasonable alternative interpretations.

\textbf{Decision Agent.}
The Decision Agent reconciles the initial hypothesis and its counter-analysis against the original branch evidence:
\begin{equation}
D_b =
\mathcal{A}_{b}^{\mathrm{dec}}(E_b,H_b,Q_b).
\end{equation}
The resulting $D_b$ contains the branch-level risk judgment together with its principal supporting evidence, counter-evidence, contextual alternatives, and remaining uncertainty. Throughout this stage, the textual and audio-visual branches remain isolated, preventing premature cross-modal convergence before arbitration.

\subsection{Cross-Modal Evidence Arbitration}

The textual and audio-visual branches capture different aspects of depression-related information and may yield compatible or conflicting assessments. Simply averaging or voting over their predicted labels ignores the strength, context, and uncertainty of the underlying evidence. DepressionAgent therefore performs cross-modal evidence arbitration over the complete branch-level reasoning outputs.

\textbf{Cross-Modal Judgment Agent.}
After independent branch-wise deliberation, the Cross-Modal Judgment Agent receives the textual assessment $D_T$ and audio-visual assessment $D_{\mathrm{AV}}$. Rather than comparing only their predicted labels, it jointly examines their supporting evidence, counter-evidence, contextual factors, alternative explanations, and remaining uncertainty:
\begin{equation}
\left(
R_{\mathrm{cm}},
\mathcal{C}_0
\right)
=
\mathcal{A}^{\mathrm{jud}}
\left(
D_T,D_{\mathrm{AV}}
\right),
\end{equation}
where $\mathcal{A}^{\mathrm{jud}}$ denotes the Cross-Modal Judgment Agent, $R_{\mathrm{cm}}\in\{\mathrm{consistent},\mathrm{conflict}\}$ characterizes the relationship between the two branch assessments, and $\mathcal{C}_0$ records the evidence comparison and arbitration rationale.

When the two branches are consistent, the agent further verifies whether their conclusions are supported by compatible evidence rather than merely exhibiting superficial label agreement. Complementary self-reported and behavioral evidence is then jointly considered to form the initial multimodal prediction:
\begin{equation}
\hat{Y}_{\mathrm{initial}}
=
\mathcal{A}^{\mathrm{risk}}
\left(
D_T,D_{\mathrm{AV}},\mathcal{C}_0
\right).
\end{equation}

\textbf{Conflict Reflection.}
When the textual and audio-visual branches provide conflicting assessments, DepressionAgent does not immediately resolve the disagreement through voting or fixed modality weighting. Instead, Conflict Reflection examines where the disagreement originates, including potentially different interpretations of ambiguous evidence, overlooked counter-evidence, or insufficient contextual consideration:
\begin{equation}
F_{\mathrm{conf}}
=
\mathcal{A}^{\mathrm{conf}}
\left(
D_T,D_{\mathrm{AV}},\mathcal{C}_0
\right),
\end{equation}
where $F_{\mathrm{conf}}$ summarizes the identified source of disagreement and the evidence requiring further examination.

The reflection is returned to the branch-wise deliberation stage so that the relevant evidence and interpretation can be reconsidered before cross-modal judgment is performed again. Conflict Reflection therefore acts as a feedback mechanism for revising the reasoning that produced the disagreement, rather than directly overriding either branch decision. After the disagreement has been reconsidered, the Cross-Modal Judgment Agent produces $\hat{Y}_{\mathrm{initial}}$ and the corresponding arbitration record $\mathcal{C}_0$.

\subsection{Risk-Sensitive Textual Reassessment}

After branch-wise deliberation and cross-modal evidence arbitration, DepressionAgent obtains an initial multimodal risk assessment. A particular failure mode may nevertheless remain: potentially important subjective experiences can be expressed explicitly in text without correspondingly salient acoustic or visual manifestations. In such cases, an initially low-risk multimodal decision may underemphasize self-reported risk evidence. DepressionAgent therefore performs an additional textual second-opinion assessment for every initially low-risk prediction.

\textbf{Risk Reflection.}
Risk Reflection serves as an asymmetric verification mechanism for low-risk decisions. Rather than repeating the complete multimodal reasoning pipeline, it redirects attention to the original transcript, where subjective symptoms and internal experiences are expressed most directly. Its purpose is to determine whether potentially important textual evidence was overlooked, underemphasized, or outweighed by ambiguous behavioral observations during the initial multimodal assessment.

The reassessment trigger is defined as
\begin{equation}
z =
\mathbb{I}
\left[
\hat{Y}_{\mathrm{initial}} = 0
\right],
\qquad
z \in \{0,1\},
\end{equation}
where $\mathbb{I}[\cdot]$ denotes the indicator function. Thus, initially high-risk predictions are retained directly, whereas every initially low-risk prediction enters the following textual second-opinion pathway.

\textbf{Text Review Agent.}
Once Risk Reflection is triggered, an independent Text Review Agent re-reads the complete original transcript and specifically searches for risk-relevant information that may have been insufficiently considered during the first-pass reasoning. It does not receive the previous textual branch decision, thereby reducing anchoring to the initial textual interpretation:
\begin{equation}
R_T =
\mathcal{A}^{\mathrm{rev}}(T),
\end{equation}
where $\mathcal{A}^{\mathrm{rev}}$ denotes the Text Review Agent and $R_T$ contains the newly reviewed textual evidence together with its contextual support.

\textbf{Re-evaluate Agent.}
The Re-evaluate Agent subsequently conducts an independent second-opinion assessment based on the complete transcript and reviewed evidence $R_T$:
\begin{equation}
\left(
\hat{Y}_{R},
\mathcal{C}_{R}
\right)
=
\mathcal{A}^{\mathrm{reev}}
\left(
T,R_T
\right),
\end{equation}
where $\mathcal{A}^{\mathrm{reev}}$ denotes the Re-evaluate Agent, $\hat{Y}_{R}$ denotes its second-opinion risk assessment, and $\mathcal{C}_{R}$ records the corresponding rationale.

The final prediction and reasoning record are therefore defined as
\begin{equation}
\left(
\hat{Y},
\mathcal{C}
\right)
=
\begin{cases}
\left(
\hat{Y}_{\mathrm{initial}},
\mathcal{C}_{0}
\right),
&
\hat{Y}_{\mathrm{initial}} = 1,
\\[4pt]
\left(
\hat{Y}_{R},
\left(
\mathcal{C}_{0},R_T,\mathcal{C}_{R}
\right)
\right),
&
\hat{Y}_{\mathrm{initial}} = 0.
\end{cases}
\end{equation}

This design does not impose a global text-first decision rule. Instead, the asymmetry is conditioned on the initial multimodal outcome: initially high-risk decisions are retained, whereas initially low-risk decisions receive an independent textual second opinion. The mechanism therefore provides a risk-oriented safeguard against false negatives without repeatedly executing the complete multimodal reasoning process.

\section{Experiments}
\label{sec:experiments}

\subsection{Datasets}

\textbf{D-Vlog Dataset.}
D-Vlog is an in-the-wild audiovisual dataset for depression assessment~\cite{yoon2022dvlog}. It contains 961 English YouTube vlogs from 816 speakers, totaling approximately 160 hours, including 555 depression and 406 non-depression samples, with access limited to 885 videos due to privacy issues. The videos are recorded in unconstrained real-world environments and contain substantial variation in speaking style, facial behavior, background conditions, and recording quality.

\textbf{LMVD Dataset.}
LMVD is a large-scale multimodal vlog dataset for depression assessment in real-world scenarios~\cite{he2026lmvd}. It contains 1,823 videos from 1,475 participants, totaling approximately 214 hours, with 908 depression and 915 non-depression samples. The videos were collected from multiple social media platforms, resulting in substantial diversity in participants, recording environments, and content.

\subsection{Implementation Details}
\label{sec:experimental_setup}

\textbf{Evaluation Protocol.}
All experiments are conducted in a depression-specific training-free zero-shot setting. DepressionAgent does not use benchmark labels for parameter updating, prompt optimization, in-context demonstrations, or model selection. For deterministic evaluation, temperature was set to 0 for all foundation-model inference unless otherwise specified.

We use two evaluation protocols for different purposes. For comparison with existing methods, DepressionAgent is evaluated on the same benchmark test partitions used for state-of-the-art (SOTA) comparison, ensuring that all reported methods are compared on the same evaluation samples, the full-dataset results in Tables II–V and VII are not used for SOTA comparison. For internal analyses on D-Vlog, including functional ablation, sub-agent ablation, reflection analysis, modality contribution, explicit-language masking, and foundation-model robustness, we merge the original training, validation, and test partitions and evaluate the fixed training-free framework on all samples. Since no benchmark labels are used to train or adapt DepressionAgent, this enlarged evaluation pool allows the behavior of the proposed reasoning mechanisms to be examined on a larger and more diverse sample set. Results obtained under this full-dataset analysis protocol are not used for direct SOTA comparison. In all experiments, ground-truth labels are accessed only after inference for metric computation.

Following previous studies, we report Accuracy (Acc.), Precision (Prec.), Recall (Rec.), and F1-score (F1), treating the risk-positive category as the positive class.

\textbf{Foundation Model.}
The default implementation adopts \textbf{Qwen3-Omni-30B-A3B-Instruct}. GPT-5.5 and Gemini~3.1~Pro are additionally used in Sec.~\ref{sec:foundation_models} to evaluate whether the proposed reasoning framework remains effective across different foundation models.

\textbf{Visual Processing.}
For each video, 64 frames are uniformly sampled along the temporal axis following the default input convention of the multimodal backbone~\cite{xu2025qwen3}. The frames are resized and normalized on-the-fly by the model processor and directly processed by Qwen3-Omni-30B-A3B-Instruct.

\textbf{Acoustic Processing.}
The audio track is extracted from each video and converted to a 16\,kHz mono waveform using ffmpeg (``-ac 1'', ``-ar 16000''). The waveform is loaded with \texttt{sr=16000, mono=True} and directly processed by Qwen3-Omni-30B-A3B-Instruct without an additional acoustic feature extractor.

\textbf{Transcript and Text-Evidence Processing.}
Transcripts are generated from the audio track using \textbf{Whisper-large-v3}~\cite{radford2023robust}. The resulting transcript is then processed by Qwen3-Omni-30B-A3B-Instruct using the dedicated textual observation prompt.

Across all experiments, the agent roles, information-access constraints, and output schemas remain fixed. Intermediate outputs are automatically checked for schema completeness; malformed outputs are regenerated using the same input and reasoning objective for at most three attempts. Complete prompts and structured output formats are provided in the supplementary material.

\subsection{Ablation Study}
\label{sec:workflow_analysis}

\subsubsection{Functional Ablation}

We first evaluate the contribution of the major reasoning components using the full D-Vlog evaluation pool. Rather than removing isolated agents from the complete system, we progressively construct the reasoning process while keeping the first-stage modality-specific evidence fixed across all configurations. The \textit{Baseline} directly produces a prediction from the extracted multimodal evidence without branch-wise deliberation, cross-modal arbitration, or risk-sensitive textual reassessment.

As shown in Table~\ref{tab:functional_ablation}, the baseline achieves 69.49\% accuracy and 78.33\% F1. Introducing Branch-Wise Deliberation substantially improves performance to 83.73\% accuracy and 86.81\% F1, demonstrating the benefit of explicitly examining supporting and counter-evidence before branch-level decisions are formed. Cross-Modal Evidence Arbitration further increases accuracy to 89.60\% and F1 to 90.21\%, showing that explicit reasoning over cross-modal agreement and disagreement provides additional benefit beyond independent branch deliberation. Finally, Risk-Sensitive Textual Reassessment raises the performance to 92.43\% accuracy and 93.20\% F1. The progressive improvements indicate complementary contributions from evidence deliberation, cross-modal reflection, and risk-oriented second opinion.

\begin{table}[t]
    \centering
    \caption{
    Functional ablation of DepressionAgent on the full D-Vlog evaluation pool. Modules are progressively added from top to bottom. Values in red denote the performance gain over the preceding configuration.
    }
    \label{tab:functional_ablation}
    \setlength{\tabcolsep}{1mm}
    \renewcommand{\arraystretch}{1.15}
    \fontsize{9}{9}\selectfont

    \begin{tabular}{lcc}
        \toprule
        \textbf{Configuration} & \textbf{Acc.} & \textbf{F1} \\
        \midrule

        Baseline
        & 69.49
        & 78.33 \\

        + Branch-Wise Deliberation
        & 83.73 {\scriptsize\textcolor{red}{(+14.24)}}
        & 86.81 {\scriptsize\textcolor{red}{(+8.48)}} \\

        + Cross-Modal Evidence Arbitration
        & 89.60 {\scriptsize\textcolor{red}{(+5.87)}}
        & 90.21 {\scriptsize\textcolor{red}{(+3.40)}} \\

        \rowcolor{gray!10}
        + Risk-Sensitive Textual Reassessment
        & \textbf{92.43} {\scriptsize\textcolor{red}{(+2.83)}}
        & \textbf{93.20} {\scriptsize\textcolor{red}{(+2.99)}} \\

        \bottomrule
    \end{tabular}
\end{table}

\subsubsection{Sub-Agent Ablation}

We further examine the complementary roles of the Hypothesis Agent and Challenger Agent in Branch-Wise Deliberation. The Hypothesis Agent performs \emph{risk discovery}, aiming to comprehensively identify potentially depression-related evidence. In contrast, the Challenger Agent performs \emph{risk exclusion} by scrutinizing the initial hypothesis, searching for contradictory evidence, and considering plausible non-depressive explanations. We separately remove each agent while keeping the remaining components unchanged.

As shown in Table~\ref{tab:ablation_deliberation}, removing either agent decreases performance. Without the Hypothesis Agent, the model achieves an F1 score of 92.48\%, suggesting that systematic risk discovery helps surface subtle potentially relevant cues. Without the Challenger Agent, F1 decreases to 91.77\%, indicating that an initial risk-oriented interpretation benefits from explicit counter-analysis. Removing both agents further decreases F1 to 90.46\%. The complete support--challenge deliberation yields the highest accuracy of 92.43\% and F1 of 93.20\%, supporting the complementary roles of risk discovery and critical verification.

\begin{table}[t]
    \centering
    \caption{
    Ablation of the Hypothesis and Challenger Agents in Branch-Wise Deliberation. All other components of DepressionAgent are kept unchanged.
    }
    \label{tab:ablation_deliberation}
    \setlength{\tabcolsep}{2.0mm}
    \renewcommand{\arraystretch}{1.2}
    \fontsize{9}{9}\selectfont

    \begin{tabular}{cc|cc}
        \toprule
        \textbf{Hypothesis Agent}
        & \textbf{Challenger Agent}
        & \textbf{Acc.}
        & \textbf{F1} \\
        \midrule

        &
        & 89.59 & 90.46 \\

        \checkmark &
        & 91.18 & 91.77 \\

        & \checkmark
        & 91.40 & 92.48 \\

        \midrule
        \rowcolor{gray!10}
        \checkmark & \checkmark
        & \textbf{92.43} & \textbf{93.20} \\

        \bottomrule
    \end{tabular}
\end{table}

\subsubsection{Behavior of the Two Reflection Mechanisms}

Beyond aggregate ablation, we directly examine how Conflict Reflection and Risk Reflection modify individual predictions. As shown in Table~\ref{tab:reflection_changes}, the initial prediction after Branch-Wise Deliberation achieves 83.73\% accuracy. Conflict Reflection changes 111 erroneous predictions to correct predictions while changing 59 initially correct predictions to incorrect ones, yielding 52 net corrections and improving accuracy to 89.60\% ($+5.87$ percentage points).

Risk Reflection subsequently operates on initially low-risk decisions. It corrects 35 remaining errors while introducing 10 adverse changes, resulting in another 25 net corrections and increasing accuracy to 92.43\% ($+2.83$ percentage points). Together, the two reflection mechanisms yield 77 net corrections and an overall accuracy improvement of 8.70 percentage points. These transition statistics show that the reflection operations more frequently repair erroneous decisions than disturb correct ones, while also making their correction behavior directly inspectable.

\begin{table}[t]
    \centering
    \caption{Prediction changes introduced by the two reflection mechanisms on the full D-Vlog evaluation pool.}
    \label{tab:reflection_changes}
    \setlength{\tabcolsep}{2.2mm}
    \renewcommand{\arraystretch}{1.2}
    \fontsize{9}{9}\selectfont

    \begin{tabular}{@{}lcccc@{}}
        \toprule
        \textbf{Stage}
        & \shortstack{\textbf{Wrong}\\$\boldsymbol{\rightarrow}$\textbf{Correct}}
        & \shortstack{\textbf{Correct}\\$\boldsymbol{\rightarrow}$\textbf{Wrong}}
        & \shortstack{\textbf{Net}\\\textbf{Correction}}
        & \textbf{Acc.} \\
        \midrule

        Initial Prediction
        & -- & -- & -- & 83.73 \\

        Conflict Reflection
        & 111 & 59 & $+52$ & 89.60 \\

        Risk Reflection
        & 35 & 10 & $+25$ & \textbf{92.43} \\

        \bottomrule
    \end{tabular}
\end{table}

\subsection{Contribution of Different Modalities}

We next examine how different modalities contribute to DepressionAgent under the full D-Vlog analysis protocol.

\begin{table}[t]
\caption{Modality contribution analysis on the full D-Vlog evaluation pool.}
\label{tab:modality_ablation}
\centering
\setlength{\tabcolsep}{4mm}
\renewcommand{\arraystretch}{1.2}
\fontsize{9}{9}\selectfont
\begin{tabular}{ccc|cc}
\toprule
\textbf{Reading} & \textbf{Seeing} & \textbf{Listening}
& \textbf{Acc.} & \textbf{F1} \\
\midrule
\checkmark &            &            & 89.60 & 90.38 \\
            & \checkmark &            & 44.68 & 0.00 \\
            &            & \checkmark & 57.13 & 39.75 \\
            & \checkmark & \checkmark & 63.84 & 54.80 \\
\midrule
\rowcolor{gray!10}
\checkmark & \checkmark & \checkmark
& \textbf{92.43} & \textbf{93.20} \\
\bottomrule
\end{tabular}
\end{table}

As shown in Table~\ref{tab:modality_ablation}, text provides the strongest unimodal performance, reaching 89.60\% accuracy and 90.38\% F1. This indicates that self-reported linguistic and contextual information carries substantial predictive information on D-Vlog. Audio and video are markedly weaker when used independently, whereas combining them improves performance relative to either behavioral modality alone, supporting their complementary role within the audio-visual branch.

Importantly, the complete multimodal framework further improves accuracy to 92.43\% and F1 to 93.20\%, outperforming the text-only setting by 2.83 and 2.82 percentage points, respectively. The results therefore do not imply that nonverbal information is redundant. Rather, audio-visual evidence provides complementary behavioral information that can support, qualify, or challenge textual interpretations during evidence arbitration.

\subsection{Qualitative Analysis}
\label{sec:qualitative_analysis}

\begin{figure*}[t]
\centering
\includegraphics[width=\linewidth]{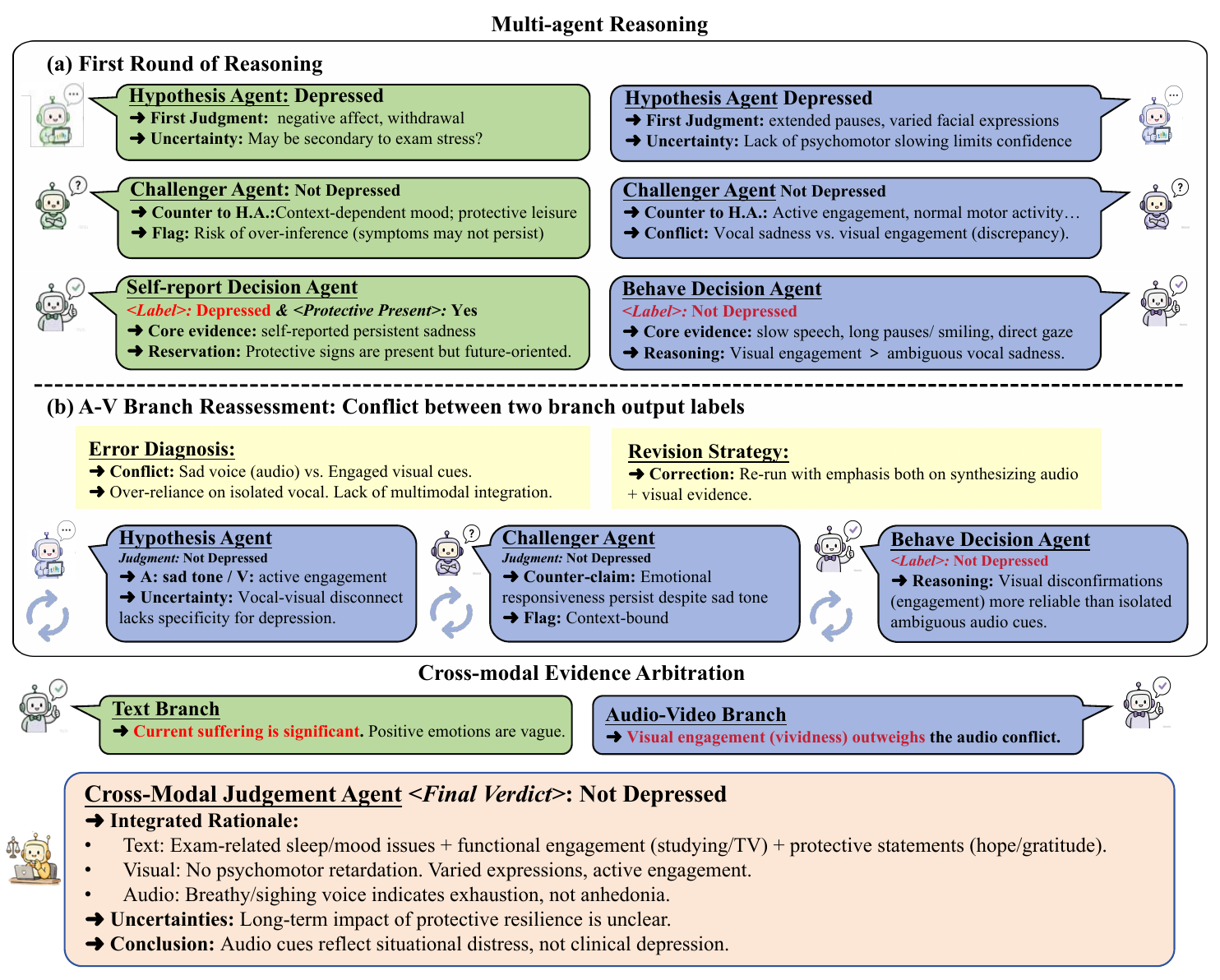}
\caption{
Representative case illustrating the evidence-level reasoning process of DepressionAgent, including branch-wise evidence deliberation, cross-modal disagreement, and reflection-based revision.
}
\label{fig:qualitative_case}
\end{figure*}

To illustrate how DepressionAgent performs evidence-level multimodal reasoning, Fig.~\ref{fig:qualitative_case} presents a representative case involving disagreement between textual and audio-visual evidence.

The text branch identifies negative mood and sleep complaints together with protective signals such as continued studying and social engagement. Although protective factors are present, the accumulation of negative self-reported experiences leads the textual branch toward a higher-risk interpretation. Meanwhile, the audio-visual branch captures potentially concerning vocal cues but also observes strong visual counter-evidence, including direct eye contact, varied facial expressions, and preserved psychomotor activity. These observations provide plausible alternatives to interpreting the vocal patterns as depression-specific evidence.

During branch-wise deliberation, the Challenger Agent explicitly questions the initial audio-visual hypothesis, noting that isolated vocal characteristics are not accompanied by sustained flat affect or behavioral disengagement. The audio-visual branch therefore revises its assessment.

Cross-modal arbitration subsequently compares the two branches rather than simply selecting one modality's prediction. It recognizes that the negative textual evidence is strongly embedded in an examination-related context, while audio-visual observations indicate preserved engagement and functioning. Considering both risk-related and counterbalancing evidence, DepressionAgent interprets the case as more consistent with situational distress than elevated depression risk.

This case illustrates the intended role of agent collaboration in DepressionAgent: intermediate interpretations can be challenged, disagreement can be explicitly exposed, and the evidence underlying that disagreement can be reconsidered before the final decision is formed.

\subsection{Robustness Across Foundation Models}
\label{sec:foundation_models}

\begin{figure*}[t]
\centering
\includegraphics[width=\linewidth]{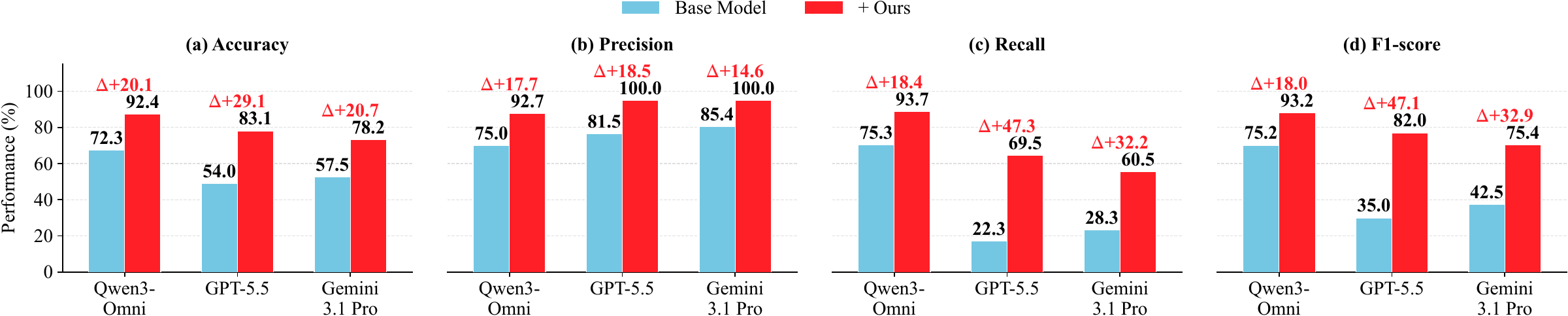}
\caption{
Comparison of foundation-model reasoning and DepressionAgent reasoning across different foundation models. For each foundation model, both settings use the same multimodal inputs, while the former performs a single-pass prediction and the latter applies the proposed evidence-centric multi-agent reasoning framework. The numbers above the bars indicate the accuracy improvement ($\Delta$ Acc.) achieved by DepressionAgent.
}
\label{fig:foundation_models}
\end{figure*}

\begin{table*}[t]
    \centering
    \caption{
        Performance comparison on the D-Vlog and LMVD datasets under matched benchmark test-set protocols.
        Results of previous methods are collected from their original publications.
        ``$-$'' denotes unavailable results.
        The best performance on each dataset is highlighted in \textbf{bold}.
        \textit{Training Free} indicates whether task-specific parameter training is required,
        and \textit{Explicit Evidence} indicates whether human-readable evidence is explicitly produced together with the prediction.
    }
    \label{tab:main_comparison}
    \setlength{\tabcolsep}{1.8mm}
    \renewcommand{\arraystretch}{1.2}
    \fontsize{9}{9}\selectfont

    \begin{tabular}{l|c|cccc|cccc|c|c}
        \toprule
        \multirow{2}{*}{\textbf{Method}}
        & \multirow{2}{*}{\textbf{Year}}
        & \multicolumn{4}{c|}{\textbf{D-Vlog}}
        & \multicolumn{4}{c|}{\textbf{LMVD}}
        & \multirow{2}{*}{\makecell{\textbf{Training}\\\textbf{Free}}}
        & \multirow{2}{*}{\makecell{\textbf{Explicit}\\\textbf{Evidence}}} \\
        \cmidrule(lr){3-6}
        \cmidrule(lr){7-10}
        &
        & \textbf{Acc.}
        & \textbf{Prec.}
        & \textbf{Rec.}
        & \textbf{F1}
        & \textbf{Acc.}
        & \textbf{Prec.}
        & \textbf{Rec.}
        & \textbf{F1}
        & & \\
        \midrule

        Depression Detector~\cite{yoon2022d}
        & 2022
        & $-$ & 65.40 & 65.57 & 63.50
        & $-$ & $-$ & $-$ & $-$
        & \ding{55} & \ding{55} \\

        TAMFN~\cite{zhou2022tamfn}
        & 2022
        & $-$ & 66.02 & 66.50 & 65.82
        & $-$ & $-$ & $-$ & $-$
        & \ding{55} & \ding{55} \\

        Attention-MIL~\cite{shangguan2023automatic}
        & 2023
        & $-$ & 67.27 & 67.77 & 66.64
        & $-$ & $-$ & $-$ & $-$
        & \ding{55} & \ding{55} \\

        CAIINET~\cite{zhou2023caiinet}
        & 2023
        & $-$ & 66.56 & 66.98 & 66.55
        & $-$ & $-$ & $-$ & $-$
        & \ding{55} & \ding{55} \\

        STST~\cite{TAO2024577}
        & 2024
        & 70.70 & 72.50 & 77.67 & 75.00
        & $-$ & $-$ & $-$ & $-$
        & \ding{55} & \ding{55} \\

        DepMSTAT~\cite{tao2024depmstat}
        & 2024
        & $-$ & 71.53 & 75.60 & 73.51
        & $-$ & $-$ & $-$ & $-$
        & \ding{55} & \ding{55} \\

        MDAVIF~\cite{ling2024mdavif}
        & 2024
        & $-$ & 74.25 & 76.00 & 75.25
        & $-$ & $-$ & $-$ & $-$
        & \ding{55} & \ding{55} \\

        Depressformer~\cite{he2024depressformer}
        & 2024
        & 65.00 & 64.00 & 54.00 & 59.00
        & $-$ & $-$ & $-$ & $-$
        & \ding{55} & \ding{55} \\

        DepMamba~\cite{10889975}
        & 2025
        & 68.87 & 68.19 & 86.99 & 76.44
        & 72.13 & 70.18 & 76.56 & 73.20
        & \ding{55} & \ding{55} \\

        CMF-Mamba~\cite{zhou2025cross}
        & 2025
        & $-$ & $-$ & $-$ & $-$
        & 74.86 & 73.20 & 78.02 & 75.53
        & \ding{55} & \ding{55} \\

        STCM-Mamba~\cite{zhou2025stcm}
        & 2025
        & 72.64 & 73.72 & 82.11 & 77.69
        & 75.41 & 72.55 & 76.68 & 76.49
        & \ding{55} & \ding{55} \\

        Xception~\cite{he2026lmvd}
        & 2026
        & $-$ & $-$ & $-$ & $-$
        & 71.38 & 71.94 & 71.38 & 71.19
        & \ding{55} & \ding{55} \\

        ViT~\cite{he2026lmvd}
        & 2026
        & $-$ & $-$ & $-$ & $-$
        & 73.03 & 73.52 & 73.03 & 72.90
        & \ding{55} & \ding{55} \\

        MSI-Net~\cite{shi2026interaction}
        & 2026
        & 70.25 & 78.91 & 69.63 & 73.91
        & 74.34 & 78.80 & 75.29 & 77.06
        & \ding{55} & \ding{55} \\

        MDDFormer~\cite{he2026lmvd}
        & 2026
        & $-$ & $-$ & $-$ & $-$
        & 76.88 & 77.02 & 76.88 & 76.85
        & \ding{55} & \ding{55} \\

        BHDD~\cite{ma2026bhdd}
        & 2026
        & $-$ & 75.98 & 75.96 & 75.94
        & 78.11 & 78.39 & 78.11 & 78.06
        & \ding{55} & \ding{55} \\

        \midrule
        \rowcolor{gray!10}

        DepressionAgent
        & --
        & \textbf{95.26}
        & \textbf{95.24}
        & \textbf{96.15}
        & \textbf{95.69}
        & \textbf{82.57}
        & \textbf{80.48}
        & \textbf{85.76}
        & \textbf{83.04}
        & \ding{51}
        & \ding{51} \\

        \bottomrule
    \end{tabular}
\end{table*}

To evaluate whether DepressionAgent depends on a specific foundation model, we instantiate the same reasoning framework using Qwen3-Omni, GPT-5.5, and Gemini~3.1~Pro. The reasoning structure and agent roles remain unchanged across backbones.

As shown in Fig.~\ref{fig:foundation_models}, DepressionAgent consistently improves performance relative to direct prediction with each corresponding foundation model. Accuracy increases by 20.1, 29.1, and 20.7 percentage points for Qwen3-Omni, GPT-5.5, and Gemini~3.1~Pro, respectively. Similar improvements are observed in F1-score, with gains of 18.0, 47.1, and 32.9 percentage points.

The improvement is particularly pronounced for models with low recall under direct prediction. For GPT-5.5 and Gemini~3.1~Pro, recall increases from 22.3\% to 69.5\% and from 28.3\% to 60.5\%, respectively, while precision remains high. These consistent gains suggest that the benefit of evidence extraction, deliberation, arbitration, and reflection is not restricted to a single foundation model, although the absolute decision behavior remains backbone dependent.

\subsection{Comparison with State-of-the-Art Methods}

Table~\ref{tab:main_comparison} compares DepressionAgent with representative multimodal depression assessment methods on D-Vlog and LMVD. In contrast to the full-dataset protocol used for the internal analyses above, this comparison evaluates DepressionAgent on the same benchmark test partitions used for SOTA comparison.

Under the matched evaluation protocol, DepressionAgent achieves competitive performance across both benchmarks without depression-specific parameter training. In addition to the final prediction, DepressionAgent explicitly produces human-readable multimodal evidence and intermediate reasoning records. Most existing feature-fusion methods, in contrast, rely on supervised benchmark optimization and do not provide explicit evidence records as part of their standard prediction output.

These results show that evidence-centric agentic reasoning can provide a competitive alternative to conventional feature-level fusion. The comparison should nevertheless be interpreted in light of the different learning paradigms: DepressionAgent performs zero-shot inference without depression-specific parameter optimization, whereas the compared approaches are predominantly supervised models trained using benchmark annotations.

\section{Discussion}

\subsection{Analysis of Explicit Language Reliance}

The strong performance of the textual modality raises a potential concern that DepressionAgent may rely primarily on explicit depression-related expressions, such as direct self-reports or diagnostic terms, as lexical shortcuts. To examine this possibility, we conduct two targeted masking experiments under the same full-D-Vlog analysis protocol used for the ablation studies, while keeping the audio and video inputs unchanged. Word-level masking removes explicit depression-related terms, whereas sentence-level masking removes the complete sentences containing such terms, thereby eliminating both the keywords and their immediate semantic context.

As shown in Table~\ref{tab:language_masking}, word-level masking causes a moderate decrease in accuracy from 92.43\% to 90.17\% and in F1-score from 93.20\% to 90.93\%. Removing the complete sentences produces a larger degradation, reducing accuracy to 86.89\% and F1 to 87.47\%. The decrease is mainly reflected in recall, which drops from 93.67\% to 88.98\% and 82.65\%, respectively, while precision remains nearly unchanged.

These results show that explicit depression-related language provides useful evidence, particularly for identifying risk-positive cases, but does not fully account for the performance of DepressionAgent. Even after the corresponding sentences are removed, the framework retains substantial predictive capability, suggesting that its decisions also draw on broader linguistic context and complementary audio-visual behavioral information rather than relying solely on direct diagnostic keywords.

\begin{table}[t]
    \centering
    \caption{
    Effect of masking explicit depression-related language on D-Vlog.
    Word-level masking removes depression-related terms, while sentence-level
    masking removes the complete sentences containing these terms.
    }
    \label{tab:language_masking}
    \setlength{\tabcolsep}{2.0mm}
    \renewcommand{\arraystretch}{1.2}
    \fontsize{9}{9}\selectfont

    \begin{tabular}{@{}lcccc@{}}
        \toprule
        \textbf{Text Setting}
        & \textbf{Acc.}
        & \textbf{Prec.}
        & \textbf{Rec.}
        & \textbf{F1} \\
        \midrule

        Original
        & \textbf{92.43}
        & 92.73
        & \textbf{93.67}
        & \textbf{93.20} \\

        Word-level Masking
        & 90.17
        & \textbf{92.96}
        & 88.98
        & 90.93 \\

        Sentence-level Masking
        & 86.89
        & 92.89
        & 82.65
        & 87.47 \\

        \bottomrule
    \end{tabular}
\end{table}

\subsection{Human Evaluation of Evidence and Reasoning}
\label{sec:human_evaluation}

The existence of natural-language intermediate outputs does not by itself establish that the generated evidence and reasoning are meaningful or useful for human review. We therefore conduct a clinician evaluation to assess the plausibility and potential utility of DepressionAgent's intermediate evidence and reasoning.

\textbf{Evaluation Protocol.}
Five clinicians with experience in mental health assessment, including two junior psychiatrists and three senior psychiatrists, independently reviewed 30 D-Vlog samples selected to cover both risk labels and different patterns of multimodal consistency. Reviewers were provided with the model's intermediate outputs, including modality-specific observations, branch-wise deliberation, cross-modal arbitration, and Risk-Sensitive Textual Reassessment when applicable, while the ground-truth labels were hidden.

The evaluation contains two complementary parts. First, each sample is rated on a five-point Likert scale along three dimensions: \textit{Evidence Relevance}, measuring whether the extracted evidence is appropriately grounded in the presented case; \textit{Reasoning Coherence}, measuring whether the reasoning provides a logically coherent transition from observations to judgment; and \textit{Clinical Utility}, measuring whether the generated evidence and reasoning can facilitate human case review. Second, clinicians assess three structured reasoning operations: evidence conflict identification, cross-modal conflict identification, and Risk-Sensitive Textual Reassessment, using binary reasonable/unreasonable judgments. In the main text, we focus on Evidence Relevance and Reasoning Coherence because these dimensions directly assess the quality of the evidence-grounded reasoning process. For these operations, we report the \textbf{Majority Reasonable Rate}, defined as the proportion of evaluated cases for which at least half of the participating clinicians consider the corresponding operation reasonable. Inter-rater agreement is additionally measured using intraclass correlation coefficient (ICC) for the Likert ratings and Fleiss' $\kappa$ for the binary judgments.

\textbf{Evaluation Results.}
The mean Likert scores for Evidence Relevance, Reasoning Coherence, and Clinical Utility are 3.65, 3.71, and 3.21, respectively. These results indicate generally favorable assessments of the generated evidence and reasoning, although the perceived clinical utility is comparatively more moderate.

\begin{figure*}[t]
\centering
\includegraphics[width=\linewidth]{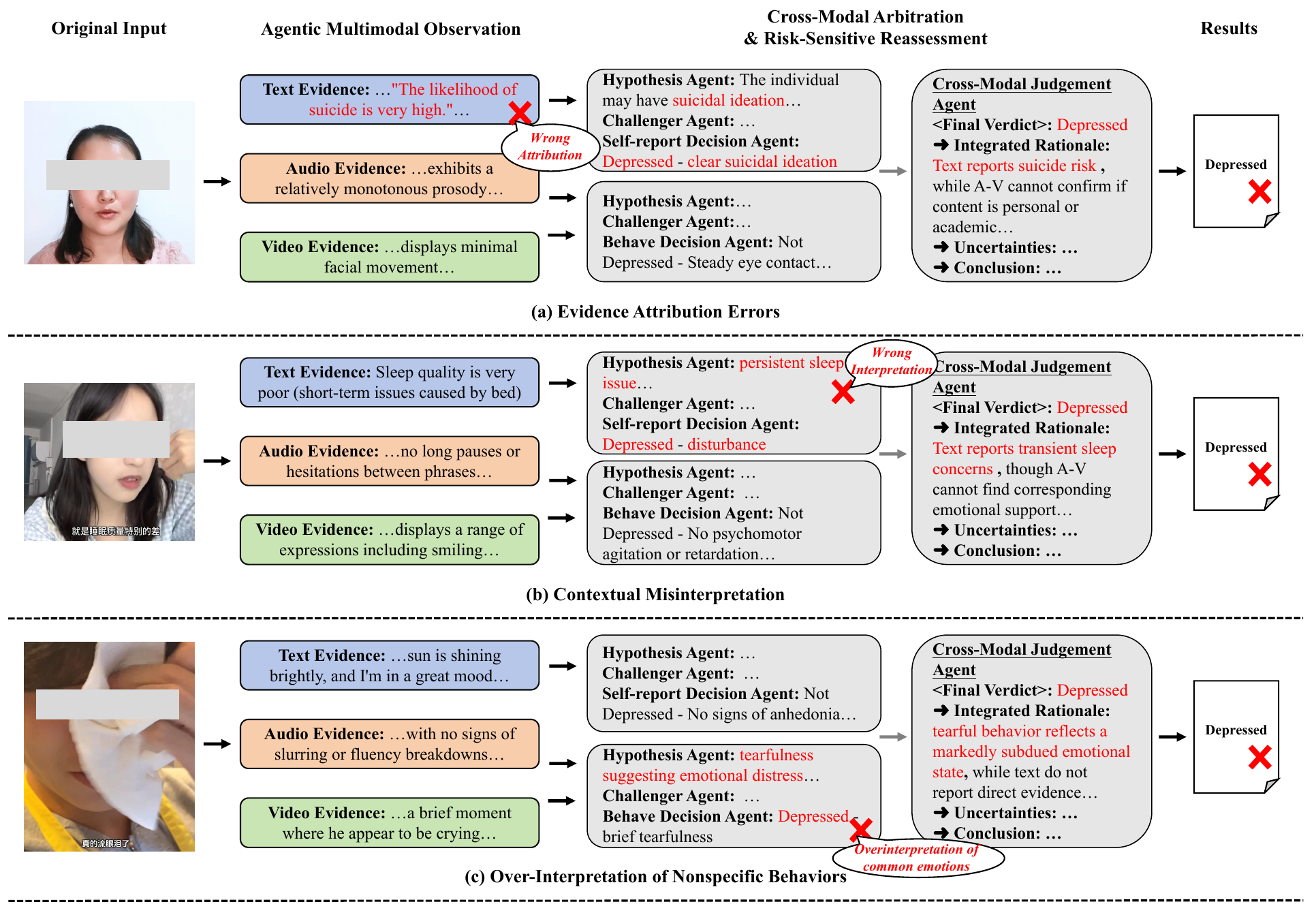}
\caption{
Illustrative examples of three representative failure modes identified by clinicians. (a) Evidence attribution error: the model attributes a third-person educational statement about suicide risk to the participant's own experience. (b) Contextual misinterpretation: a situational complaint about poor sleep caused by uncomfortable bedding is interpreted as evidence of persistent sleep disturbance. (c) Over-interpretation of non-specific behaviors: physiological tearing is interpreted as emotional distress without adequately considering alternative explanations.
}
\label{fig:failure_examples}
\end{figure*}

\begin{figure}[t]
\centering
\includegraphics[width=\columnwidth]{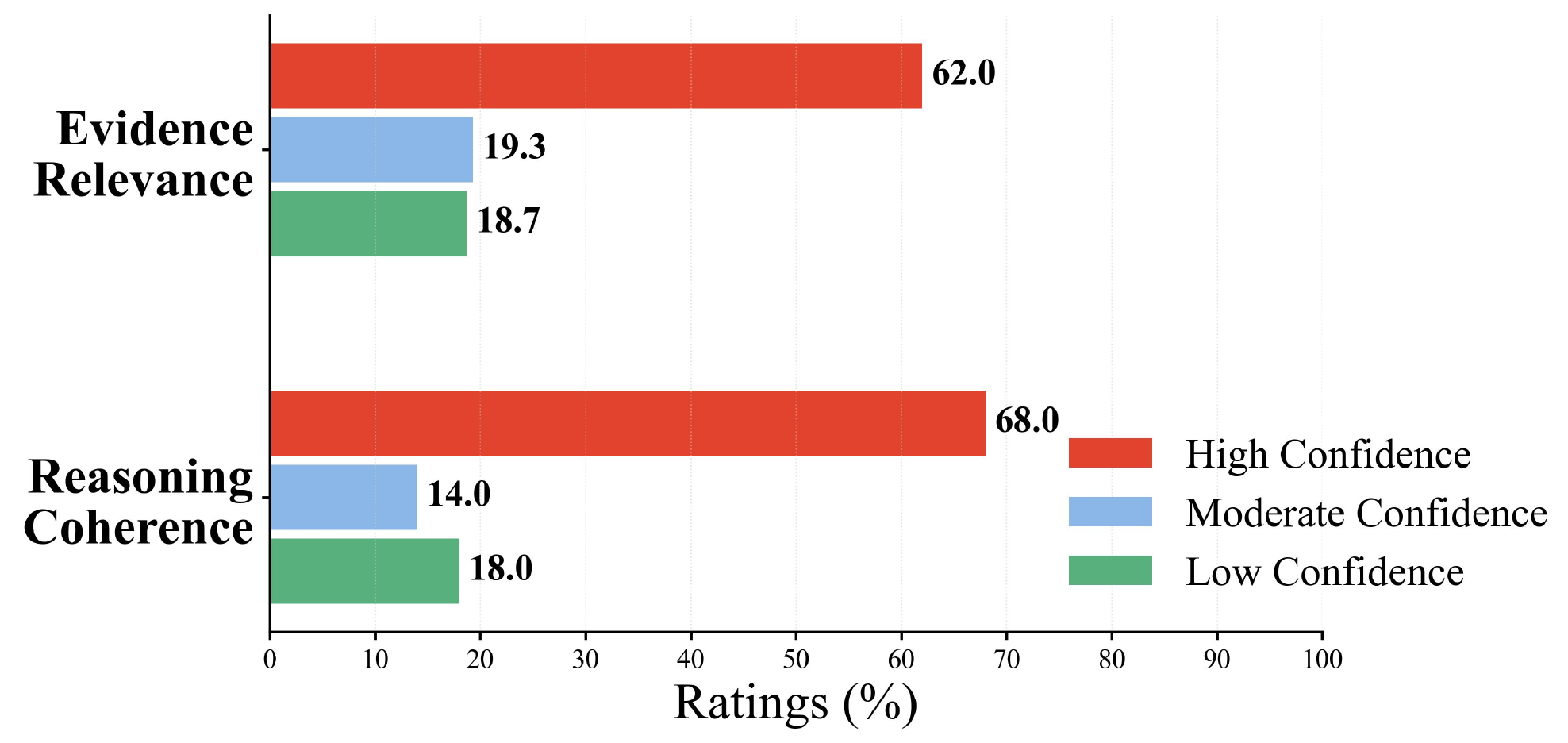}
\caption{Distribution of clinician ratings for the two reported evaluation dimensions: Evidence Relevance and Reasoning Coherence. Ratings are grouped into low, moderate, and high confidence categories.}
\label{fig:doctor_evaluation_grouped_confidence}
\end{figure}

As shown in Figure~\ref{fig:doctor_evaluation_grouped_confidence}, the majority of ratings fall into the high-confidence category. For Evidence Relevance and Reasoning Coherence, 62.0\% and 68.0\% of ratings are high-confidence, respectively. Combining moderate and high-confidence ratings, 81.3\% of Evidence Relevance ratings and 82.0\% of Reasoning Coherence ratings reach or exceed the scale mid-point, indicating that clinicians generally found the extracted evidence and the associated reasoning process to be relevant and coherent.





\begin{table}[t]
\centering
\caption{Expert assessment of structured reasoning operations.}
\label{tab:structured_operations}
\setlength{\tabcolsep}{1.0mm}
\renewcommand{\arraystretch}{1.2}
\resizebox{\columnwidth}{!}{%
\begin{tabular}{lcc}
\toprule
\textbf{Reasoning Operation}
& \textbf{Majority Reasonable}
& \textbf{Fleiss $\kappa$} \\
\midrule
Cross-modal Conflict Identification
& 90.0\% & 0.161 \\
Risk-Sensitive Textual Reassessment
& 91.7\% & 0.180 \\
\bottomrule
\end{tabular}%
}
\end{table}

Table~\ref{tab:structured_operations} further reports the two operations most directly related to cross-modal evidence adjudication and risk-sensitive reassessment. Cross-modal Conflict Identification achieves a 90.0\% majority reasonable rate (9/10 cases), while Risk-Sensitive Textual Reassessment reaches 91.7\% (11/12 cases), indicating that these two operations are considered reasonable in most evaluated cases.

Inter-rater agreement is relatively limited. Fleiss' $\kappa$ for the two reported structured reasoning operations were 0.161 and 0.180, while ICC for the two reported Likert dimensions were 0.067 and 0.289. These results indicate substantial variability among individual reviewers and suggest that the majority-reasonable rates should be interpreted as descriptive evidence of perceived plausibility rather than strong expert consensus. Accordingly, the clinician evaluation assesses the interpretability and plausibility of the generated reasoning rather than establishing clinical validity.

Beyond the quantitative ratings, clinicians' written feedback provides additional insight into how the reasoning process succeeds or fails. In several cases, reviewers noted that Risk Reflection recovered information that had been insufficiently considered in the initial multimodal assessment. The same feedback also revealed several recurring failure patterns, which we analyze next.

\subsection{Clinician-Guided Error Analysis}
\label{sec:error_analysis}

Clinicians' written feedback reveals three recurring failure modes in the evidence-deliberation process: evidence attribution errors, contextual misinterpretation, and over-interpretation of non-specific behaviors. These cases illustrate how errors introduced during evidence extraction or interpretation may propagate through subsequent reasoning.

\textbf{Evidence Attribution Errors.}
The model may incorrectly attribute evidence to its source. For example, third-person descriptions of depression were occasionally treated as the participant's self-disclosure, while background music lyrics were interpreted as participant speech. Such errors reflect limitations in speaker attribution and narrative-perspective tracking, and may contaminate subsequent reasoning even when the reasoning itself remains internally coherent.

\textbf{Contextual Misinterpretation.}
The model may misinterpret sarcasm, humor, or situational remarks as genuine depression-related expressions. For example, poor sleep caused by uncomfortable bedding was interpreted as persistent sleep disturbance. This failure reflects limitations in pragmatic language understanding and contextual interpretation.

\textbf{Over-Interpretation of Non-specific Behaviors.}
The audio-visual branch may assign excessive risk relevance to behaviors that admit alternative explanations. Physiological tearing, for instance, may be interpreted as emotional distress, while a naturally quiet speaking style may be interpreted as psychomotor slowing. This reflects a fundamental challenge in multimodal depression assessment: observable behavioral cues are often non-specific.

Figure~\ref{fig:failure_examples} illustrates representative examples of these failure modes. Clinicians also identified less frequent problems involving inappropriate evidence weighting and incomplete reassessment. Overall, explicit evidence deliberation makes such failures directly inspectable, but also exposes a key bottleneck: reasoning quality depends on whether the underlying evidence is factual, correctly attributed, properly contextualized, and appropriately weighted. Improving speaker attribution, pragmatic understanding, multimodal source separation, and context-aware evidence weighting therefore represents an important direction for future work.

\section{Conclusion}
\label{sec:conclusion}

This work presented \textbf{DepressionAgent}, an evidence-centric and reflection-driven framework for multimodal depression risk assessment. Instead of relying on latent feature fusion, DepressionAgent explicitly organizes textual, acoustic, and visual evidence through modality-specific observation, branch-wise deliberation, cross-modal arbitration, and reflection. Without depression-specific supervised training or parameter fine-tuning, it achieves competitive performance on two public benchmarks, while ablations, cross-model experiments, qualitative analysis, and clinician evaluation support the effectiveness and inspectability of the proposed reasoning process.

The current framework remains limited by errors in multimodal perception, evidence attribution, and contextual interpretation, while Risk Reflection introduces additional inference cost and textual reliance. Future work will focus on more reliable evidence grounding, adaptive reflection, uncertainty-aware reasoning, and prospective evaluation in real-world mental health settings.

\bibliographystyle{IEEEtran}
\bibliography{Ref}

\vfill

\newpage

\section*{Appendix}

The appendix provides the complete prompts used for all agents in the proposed framework. These prompts specify the role definition, task instructions, reasoning focus, and expected output format of each agent, enabling a more transparent understanding of the multi-agent reasoning process and facilitating reproducibility of our method.

\begin{figure*}[b]
\centering
\begin{promptbox}{Box A1: Prompt for Video Observation}
\small

\textbf{Role.}
You are operating inside a depression-risk assessment pipeline. Your role: Visual Behavioral Evidence Extractor.

\vspace{0.4em}
\textbf{Hard Rules.}
\begin{itemize}[leftmargin=1.2em, itemsep=0.2em]
    \item Do NOT diagnose depression. Do NOT estimate risk. Do NOT infer thoughts, emotions, motives, or causes.
    \item Describe only directly observable behavior.
    \item Behavioral dynamics are more important than static descriptions.
    \item Do NOT primarily summarize average behavior unless no meaningful variation is observed throughout the clip.
    \item If a localized but clear behavioral change occurs, report it even if it occupies only a small portion of the clip.
\end{itemize}

\vspace{0.4em}
\textbf{Prioritization.}
\begin{itemize}[leftmargin=1.2em, itemsep=0.2em]
    \item Prioritize: transitions, variability, reactivity, recovery, consistency vs rigidity
\end{itemize}

\vspace{0.4em}
\textbf{Evidence Hierarchy (highest priority first).}
\begin{itemize}[leftmargin=1.2em, itemsep=0.2em]
    \item Behavioral transitions
    \item Behavioral variability or rigidity
    \item Sustained baseline patterns
    \item Average-state descriptions
\end{itemize}

\vspace{0.4em}
\textbf{Focus on.}
\begin{itemize}[leftmargin=1.2em, itemsep=0.2em]
    \item facial expressivity and affect variability
    \item gaze engagement and attentional shifts
    \item psychomotor activity and movement dynamics
    \item posture and body-orientation changes
    \item self-adaptor behaviors if clearly observable
    \item visible response timing patterns
\end{itemize}

\vspace{0.4em}
\textbf{Output Format.}
Use the EXACT format:
\[
\left\{
\begin{array}{l}
\texttt{Major Behavioral Event 1: ...,}\\
\texttt{Major Behavioral Event 2: ...,}\\
\texttt{Major Behavioral Event 3: ...,}\\
\texttt{Stable Baseline Pattern: ...,}\\
\texttt{Behavioral Summary: ...,}\\
\texttt{Limitations: ...,}\\
\end{array}
\right\}
\]

\end{promptbox}

\label{fig:prompt_video_observation}
\end{figure*}

\begin{figure*}[!t]
\centering
\begin{promptbox}{Box A2: Prompt for Text Observation}
\small

\textbf{Role.}
You are a TEXT EVIDENCE EXTRACTOR in a depression-risk assessment pipeline.

\vspace{0.4em}
\textbf{Hard Rules.}
\begin{itemize}[leftmargin=1.2em, itemsep=0.2em]
    \item Do NOT diagnose depression. Do NOT estimate risk. Do NOT assign labels, severity, or scores.
    \item Extract evidence only. Evidence MUST be grounded in the speaker's own words.
    \item Do NOT invent context, motives, emotions, or explanations.
    \item Preserve negation, uncertainty, frequency, and time references whenever present.
\end{itemize}

\vspace{0.4em}
\textbf{Evidence Extraction.}
\begin{itemize}[leftmargin=1.2em, itemsep=0.2em]
    \item Prefer verbatim quotes ($<=$25 words).
    \item If a quote is too long, use the shortest representative excerpt.
    \item If paraphrasing is unavoidable, use only a short label ($<$8 words) after a representative quote.
    \item Do NOT add new information.
    \item Prefer high-information evidence. Deduplicate aggressively.
    \item Do NOT include multiple quotes expressing essentially the same idea unless they add distinct information.
\end{itemize}

\vspace{0.4em}
\textbf{Extract ONLY Explicitly Stated Evidence Related.}
\begin{itemize}[leftmargin=1.2em, itemsep=0.2em]
    \item \texttt{Direct self-reported internal states}: Examples: mood, sadness, hopelessness, guilt/self-blame, emotional detachment, motivation, energy/fatigue, loneliness
    \item \texttt{Behavioral / functional patterns}: Examples: social withdrawal, reduced activity, avoidance, changes in routine, work/school functioning, daily functioning, help-seeking behavior
    \item \texttt{Cognitive / framing patterns}: Examples: absolutist language, negative self-evaluation, pessimistic framing, perceived burden, helplessness, self-critical statements
    \item \texttt{Protective / counterbalancing evidence}: Include ONLY explicitly stated evidence of: future plans, goals,purpose, enjoyment, humor, social support, motivation, coping efforts, recovery-oriented actions,help-seeking intentions
    \item \texttt{Context / qualifications}: Examples: temporary stressors, situational explanations, contradictory statements, uncertainty, limited information, contextual qualifiers
\end{itemize}

\vspace{0.4em}
\textbf{Output Rules.}
\begin{itemize}[leftmargin=1.2em, itemsep=0.2em]
    \item Maximum 6 bullets per section.
    \item Each bullet $<=$25 words.
    \item Use quotes "..." whenever possible.
    \item Use labels $<...>$ only when necessary.
    \item If no evidence exists for a section, write: **None identified**
\end{itemize}

\vspace{0.4em}
\textbf{Output Format.}
Use the EXACT format:
\[
\left\{
\begin{array}{l}
\texttt{Direct Self-Reported States: ...,}\\
\texttt{Behavioral / Functional Evidence: ...,}\\
\texttt{Cognitive / Framing Patterns: ...,}\\
\texttt{Protective / Counterbalancing Evidence: ...,}\\
\texttt{Context / Qualifications: ...,}\\
\end{array}
\right\}
\]

\end{promptbox}

\label{fig:prompt_text_observation}
\end{figure*}

\begin{figure*}[t]
\centering
\begin{promptbox}{Box A3: Prompt for Audio Observation}
\small

\textbf{Role.}
You are operating inside a depression-risk assessment pipeline. Your role: Audio Behavioral Evidence Extractor.

\vspace{0.4em}
\textbf{Hard Rules.}
\begin{itemize}[leftmargin=1.2em, itemsep=0.2em]
    \item Do NOT diagnose depression. Do NOT assess risk. Do NOT infer internal emotional states.
    \item Describe only directly observable vocal behavior.
    \item Longitudinal vocal behavior is more important than average acoustic characteristics.
    \item Do NOT primarily summarize average pitch, average energy, average emotional tone, or average speaking style.
    \item If a localized but clear vocal change occurs, report it even if it occupies only a small portion of the recording.
\end{itemize}

\vspace{0.4em}
\textbf{Prioritization.}
\begin{itemize}[leftmargin=1.2em, itemsep=0.2em]
    \item Prioritize: variability, transitions, adaptation, timing dynamics, consistency vs rigidity
\end{itemize}

\vspace{0.4em}
\textbf{Evidence Hierarchy (highest priority first).}
\begin{itemize}[leftmargin=1.2em, itemsep=0.2em]
    \item Vocal transitions
    \item Vocal variability or rigidity
    \item Sustained baseline patterns
    \item Average-state descriptions
\end{itemize}

\vspace{0.4em}
\textbf{Observe.}
\begin{itemize}[leftmargin=1.2em, itemsep=0.2em]
    \item prosodic variability
    \item speech timing dynamics
    \item pause and hesitation patterns
    \item pacing and fluency changes
    \item vocal energy dynamics
    \item observable elaboration patterns
    \item conversational momentum
    \item vocal consistency versus adaptation
\end{itemize}

\vspace{0.4em}
\textbf{Output Format.}
Use the EXACT format:
\[
\left\{
\begin{array}{l}
\texttt{Major Vocal Event 1: ...,}\\
\texttt{Major Vocal Event 2: ...,}\\
\texttt{Major Vocal Event 3: ...,}\\
\texttt{Stable Baseline Pattern: ...,}\\
\texttt{Vocal Summary: ...,}\\
\texttt{Limitations: ...,}\\
\end{array}
\right\}
\]

\end{promptbox}

\label{fig:prompt_audio_observation}
\end{figure*}

\begin{figure*}[t]
\centering
\begin{promptbox}{Box A4: Prompt for Text Hypothesis Agent}
\small

\textbf{Role.}
You are the Text Liberal Analyst (T-LA), a Senior Psychiatrist focusing on depression-risk signals in transcript evidence. You will ONLY see extracted transcript evidence (already compressed/structured from the transcript).

\vspace{0.4em}
\textbf{Goal.}
Maximize sensitivity (do not miss risk), but stay grounded. Do not invent quotes or facts.

\vspace{0.4em}
\textbf{Output Rules.}
\begin{itemize}[leftmargin=1.2em, itemsep=0.2em]
    \item Write in English.
    \item Do NOT use markdown/headings/tables.
    \item Use the exact template below.
    \item Keep the whole output under 1400 characters.
\end{itemize}

\vspace{0.4em}
\textbf{Output Format.}
Use the EXACT format:
\[
\left\{
\begin{array}{l}
\texttt{Text Risk Hypothesis (supported by evidence): ...,}\\
\texttt{Potential Depression-Specific Indicators: ...,}\\
\texttt{Potential Non-Depression / Situational Explanations: ...,}\\
\texttt{Preliminary Judgment: $<$Depressed$|$Not Depressed$>$, ...}\\
\texttt{Key Uncertainties: ...,}\\
\end{array}
\right\}
\]

\end{promptbox}

\label{fig:prompt_text_hypothesis_agent}
\end{figure*}

\begin{figure*}[t]
\centering
\begin{promptbox}{Box A5: Prompt for Text Challenger Agent}
\small

\textbf{Role.}
You are the Text Conservative Critic (T-CC), a Senior Psychiatrist focusing on false-positive control using transcript evidence. You will ONLY see extracted transcript evidence (already compressed/structured from the transcript) and the T-LA output.

\vspace{0.4em}
\textbf{Goal.}
Maximize specificity while being fair and evidence-grounded.

\vspace{0.4em}
\textbf{Anti-Anchoring Rule.}
Treat T-LA output as a hypothesis, not an authority. If T-LA makes a claim not supported by the provided evidence, mark it as unsupported/over-inference flags.

\vspace{0.4em}
\textbf{Output Rules.}
\begin{itemize}[leftmargin=1.2em, itemsep=0.2em]
    \item Write in English.
    \item Do NOT use markdown/headings/tables.
    \item Use the exact template below.
    \item Keep the whole output under 1400 characters.
\end{itemize}

\vspace{0.4em}
\textbf{Output Format.}
Use the EXACT format:
\[
\left\{
\begin{array}{l}
\texttt{Primary Counterpoints to T-LA: ...,}\\
\texttt{Protective / Counterbalancing Evidence (if present in evidence): ...,}\\
\texttt{Unsupported / Over-inference Flags: ...,}\\
\texttt{Preliminary Judgment: $<$Depressed$|$Not Depressed$>$, ...}\\
\texttt{Key Uncertainties: ...,}\\
\end{array}
\right\}
\]

\end{promptbox}

\label{fig:prompt_text_challenger_agent}
\end{figure*}

\begin{figure*}[t]
\centering
\begin{promptbox}{Box A6: Prompt for Self-report Decision Agent}
\small

\textbf{Role.}
You are the Transcript Arbiter (TA) in a depression-risk assessment pipeline. Goal: compress transcript evidence into a compact, balanced, and directly traceable decision card.

\vspace{0.4em}
\textbf{Inputs.}
\begin{itemize}[leftmargin=1.2em, itemsep=0.2em]
    \item Extracted transcript evidence (already compressed/structured from the transcript)
    \item T-LA output
    \item T-CC output
\end{itemize}

\vspace{0.4em}
\textbf{Decision Policy.}
\begin{itemize}[leftmargin=1.2em, itemsep=0.2em]
    \item Weight statements by verifiability and specificity (do not default to whoever sounds more authoritative).
    \item Do NOT invent quotes or facts. Integrate all text-evidence aspects when present.
    \item Calibrate for context vs pervasiveness: if distress is mainly tied to an identifiable situational stressor or single life domain with preserved interests, planning/future orientation, routines, humor, social connection, or coping attempts, avoid over-pathologizing.
\end{itemize}

\vspace{0.4em}
\textbf{Anti-Narrative Tip.}
\begin{itemize}[leftmargin=1.2em, itemsep=0.2em]
    \item No "why" explanations.
    \item Use atomic, directly traceable fragments (quotes/cues) instead of sentences.
    \item Avoid causal words: because / therefore / suggests / indicates / implies.
\end{itemize}

\vspace{0.4em}
\textbf{Protective Present Definition.}
\begin{itemize}[leftmargin=1.2em, itemsep=0.2em]
    \item Yes = the evidence contains a protective/counterbalancing section with at least one concrete bullet/quote/claim (not "None identified").
    \item No = that section exists but has no concrete items or explicitly says "None identified".
    \item Unclear = the section cannot be reliably located.
\end{itemize}

\vspace{0.4em}
\textbf{Output Rules.}
\begin{itemize}[leftmargin=1.2em, itemsep=0.2em]
    \item Write in English.
    \item Do NOT use markdown/headings/tables.
    \item Use the exact template below.
    \item Keep the whole output under 900 characters.
\end{itemize}

\vspace{0.4em}
\textbf{Output Format.}
Use the EXACT format:
\[
\left\{
\begin{array}{l}
\texttt{[TEXT MODULE]}\\
\texttt{Label: $< $Depressed$ | $Not Depressed$ >$, ...}\\
\texttt{Protective Present: $< $Yes$ | $No$ | $Unclear$ >$, ...}\\
\texttt{Core Positive Evidence (protective/contrasting): ...,}\\
\texttt{Core Negative Evidence (risk-increasing): ...,}\\
\texttt{Context / Domain-Limited Notes: ...,}\\
\texttt{Key Uncertainties: ...,}\\
\end{array}
\right\}
\]

\end{promptbox}

\label{fig:prompt_self-report_decision_agent}
\end{figure*}

\begin{figure*}[t]
\centering
\begin{promptbox}{Box A7: Prompt for Audio-Video Hypothesis Agent}
\small

\textbf{Role.}
You are the Liberal Analyst (LA), a Senior Psychiatrist specializing in depression-risk assessment. You will ONLY see audio and visual observations (already extracted).

\vspace{0.4em}
\textbf{Goal.}
Maximize sensitivity (do not miss risk) while staying grounded in the observations.

\vspace{0.4em}
\textbf{Output Rules.}
\begin{itemize}[leftmargin=1.2em, itemsep=0.2em]
    \item Write in English.
    \item Do NOT use markdown/headings/tables.
    \item Do NOT include any role/task/prompt text.
    \item Keep the whole output under 1400 characters.
    \item Use the exact template below.
\end{itemize}

\vspace{0.4em}
\textbf{Output Format.}
Use the EXACT format:
\[
\left\{
\begin{array}{l}
\texttt{Salient Audio Cues: ...,}\\
\texttt{Salient Visual Cues: ...,}\\
\texttt{Preliminary Judgment: $<$Depressed$|$Not Depressed$>$, ...}\\
\texttt{Key Uncertainties: ...,}\\
\end{array}
\right\}
\]

\end{promptbox}

\label{fig:prompt_Audio-Video_hypothesis_agent}
\end{figure*}

\begin{figure*}[t]
\centering
\begin{promptbox}{Box A8: Prompt for Audio-Video Challenger Agent}
\small

\textbf{Role.}
You are the Conservative Critic (CC), a Senior Psychiatrist specializing in differential diagnosis and false-positive control. You will ONLY see audio and visual observations (already extracted) and the LA output.

\vspace{0.4em}
\textbf{Goal.}
Maximize specificity while being fair and evidence-grounded.

\vspace{0.4em}
\textbf{Anti-Anchoring Rule.}
\begin{itemize}[leftmargin=1.2em, itemsep=0.2em]
    \item Treat LA output as a hypothesis, not an authority.
    \item Your primary input is the raw observations. If LA makes a claim not supported by observations, label it as unsupported/over-inference flags.
\end{itemize}

\vspace{0.4em}
\textbf{Output Rules.}
\begin{itemize}[leftmargin=1.2em, itemsep=0.2em]
    \item Write in English.
    \item Do NOT use markdown/headings/tables.
    \item Do NOT include any role/task/prompt text.
    \item Keep the whole output under 1400 characters.
    \item Use the exact template below.
\end{itemize}

\vspace{0.4em}
\textbf{Output Format.}
Use the EXACT format:
\[
\left\{
\begin{array}{l}
\texttt{Primary Counterpoints to LA: ...,}\\
\texttt{Alternative Explanations: ...,}\\
\texttt{Unsupported / Over-inference Flags: ...,}\\
\texttt{Preliminary Judgment: $<$Depressed$|$Not Depressed$>$, ...}\\
\texttt{Key Uncertainties: ...,}\\
\end{array}
\right\}
\]

\end{promptbox}

\label{fig:prompt_Audio-Video_challenger_agent}
\end{figure*}

\begin{figure*}[t]
\centering
\begin{promptbox}{Box A9: Prompt for Behave Decision Agent}
\small

\textbf{Role.}
You are the Observational Arbiter (OA) in a depression-risk assessment pipeline. Goal: compress observable descriptors into a compact, balanced, and directly traceable decision card.

\vspace{0.4em}
\textbf{Inputs.}
\begin{itemize}[leftmargin=1.2em, itemsep=0.2em]
    \item Audio\&Visual observations (already extracted)
    \item LA output
    \item CC output
\end{itemize}

\vspace{0.4em}
\textbf{Decision Policy.}
\begin{itemize}[leftmargin=1.2em, itemsep=0.2em]
    \item Weight statements by verifiability and specificity (do not default to whoever sounds more authoritative).
    \item Use ONLY observable descriptors: what is seen/heard (e.g., pitch variability, pauses, speech rate, facial expressivity, gaze direction, posture, movement, response latency).
    \item Do NOT use clinical-style abstractions or psychiatric narrative (e.g., "depressive affect", "psychomotor retardation", "hopelessness", "low motivation", "social withdrawal") unless they are explicitly phrased as an observable behavior in the Audio\&Visual observations.
    \item Only retain claims that are directly grounded in the Audio\&Visual observations OR explicitly stated by LA/CC as grounded in observations.
    \item If a LA/CC claim is not supported by observations, list it under Unsupported/Over-inference.
\end{itemize}

\vspace{0.4em}
\textbf{Anti-Narrative Tip.}
\begin{itemize}[leftmargin=1.2em, itemsep=0.2em]
    \item No "why" explanations.
    \item Use atomic, directly traceable fragments (quotes/cues) instead of sentences.
    \item Avoid causal words: because / therefore / suggests / indicates / implies.
\end{itemize}

\vspace{0.4em}
\textbf{Output Rules.}
\begin{itemize}[leftmargin=1.2em, itemsep=0.2em]
    \item Write in English.
    \item Do NOT use markdown/headings/tables.
    \item Use the exact template below.
    \item Keep the whole output under 900 characters.
\end{itemize}

\vspace{0.4em}
\textbf{Output Format.}
Use the EXACT format:
\[
\left\{
\begin{array}{l}
\texttt{[Audio\&Visual MODULE]}\\
\texttt{Label: $<$Depressed$|$Not Depressed$>$, ...}\\
\texttt{Reliable Observable Cues: ...,}\\
\texttt{Alternative Explanations: ...,}\\
\texttt{Audio\&Visual Limitations: ...,}\\
\texttt{Unsupported / Over-inference Flags: ...,}\\
\end{array}
\right\}
\]

\end{promptbox}

\label{fig:prompt_behave_decision_agent}
\end{figure*}

\begin{figure*}[t]
\centering
\begin{promptbox}{Box A10: Prompt for Audio-Video Reassessment}
\small

\textbf{Role.}
You are a reflection controller for an Audio\&Visual-only branch.

\vspace{0.4em}
\textbf{Inputs.}
\begin{itemize}[leftmargin=1.2em, itemsep=0.2em]
    \item Audio\&Visual observations (already extracted)
    \item LA output
    \item CC output
    \item OA output
\end{itemize}

\vspace{0.4em}
\textbf{Task.}
\begin{itemize}[leftmargin=1.2em, itemsep=0.2em]
    \item Diagnose why the Audio\&Visual branch may be over/under-estimating risk.
    \item Focus on grounding and avoiding hallucinations / over-inference from sparse or non-specific Audio\&Visual cues.
    \item Provide a concrete revision strategy to rerun LA/CC/OA once.
\end{itemize}

\vspace{0.4em}
\textbf{Output Rules.}
\begin{itemize}[leftmargin=1.2em, itemsep=0.2em]
    \item Output must be a single valid JSON object (double quotes, no trailing commas).
    \item No markdown.
\end{itemize}

\vspace{0.4em}
\textbf{Output Format.}
Use the EXACT format:
\[
\left\{
\begin{array}{l}
\texttt{"conflict\_points": "...",}\\
\texttt{"error\_diagnosis": "...",}\\
\texttt{"revision\_strategy": "..."}
\end{array}
\right\}
\]

\end{promptbox}

\label{fig:prompt_Audio-Video_reassessment}
\end{figure*}

\begin{figure*}[t]
\centering
\begin{promptbox}{Box A11: Prompt for Cross-Modal Judgement Agent}
\small

\textbf{Role.}
You are the Judge Moderator (JM), a Senior Psychiatrist and methodologist.

\vspace{0.4em}
\textbf{Goal.}
Produce the final decision. Ensure the output is stable and evidence-grounded.

\vspace{0.4em}
\textbf{Inputs.}
\begin{itemize}[leftmargin=1.2em, itemsep=0.2em]
    \item TA output (text branch final decision card, including transcript evidence)
    \item OA output (Audio\&Visual branch final decision card)
\end{itemize}

\vspace{0.4em}
\textbf{Decision Policy.}
\begin{itemize}[leftmargin=1.2em, itemsep=0.2em]
    \item If TA and OA conflict, prefer the interpretation that is better supported by specific, verifiable evidence (quotes for text; observable descriptors for Audio\&Visual).
    \item When Audio\&Visual seems non-specific, explicitly consider benign alternatives (e.g., vlog style, recording fatigue, camera/lighting constraints, sparse observations).
    \item Behavioral and social context evidence may strengthen or weaken confidence in text-based conclusions even when it cannot independently establish depression.
\end{itemize}

\vspace{0.4em}
\textbf{Output Rules.}
\begin{itemize}[leftmargin=1.2em, itemsep=0.2em]
    \item Write in English.
    \item Do NOT use markdown/headings/tables.
    \item Use the exact template below.
    \item Keep the whole output under 1500 characters.
\end{itemize}

\vspace{0.4em}
\textbf{Output Format.}
Use the EXACT format:
\[
\left\{
\begin{array}{l}
\texttt{Final Verdict: $<$Depressed$|$Not Depressed$>$, ...}\\
\texttt{Cross-Modal Consistency: $<$Yes$|$Partially$|$No$>$, ...}\\
\texttt{If No: Why Audio\&Visual may be non-specific: ...,}\\
\texttt{Key Integrated Rationale (cite TA/OA items): ...,}\\
\texttt{Key Uncertainties: ...,}\\
\end{array}
\right\}
\]

\end{promptbox}

\label{fig:prompt_Cross-Modal_judgement_agent}
\end{figure*}

\begin{figure*}[t]
\centering
\begin{promptbox}{Box A12: Prompt for Text Review Agent}
\small

\textbf{Role.}
You are a Transcript Review Gate. You will receive a complete transcript. Your task is NOT to diagnose depression. Your task is NOT to estimate risk. Your task is only to determine whether the transcript contains highly specific evidence that warrants an additional review.

\vspace{0.4em}
\textbf{Assessment Dimensions.}
\begin{itemize}[leftmargin=1.2em, itemsep=0.2em]
    \item \texttt{historical\_depression}: Count ONLY if explicitly stated: Explicit self-report of being depressed. / Explicit depression diagnosis. / Explicit depression history. / Explicit statement of having experienced depression in the past. Do NOT count: General sadness. / Stress. / Anxiety alone. / Indirect implications.
    \item \texttt{suicidal\_content}: Count ONLY if explicitly stated: Explicit suicidal ideation. / Suicide attempt. / Self-harm intent. / Recurrent thoughts of death about self. / Desire to die. Do NOT count: Metaphorical expressions. / Discussion of suicide unrelated to self.
    \item \texttt{functional\_impairment}: Count ONLY if explicitly stated: Explicit impairment in work, school, relationships, self-care, or daily functioning attributed to emotional or mental difficulties. Examples: Unable to work. / Dropped out of school. / Stayed in bed most days. / Significant social withdrawal affecting daily life. Do NOT count: Temporary inconvenience. / Normal stress reactions.
    \item \texttt{persistent\_negative\_self\_view}: Count ONLY if explicitly stated: Repeated and generalized negative self-evaluation. / Worthlessness. / Self-hatred. / Global negative identity statements. Examples: I am worthless. / I hate myself. / I am a failure. / I am nothing. Do NOT count: Situation-specific criticism. / A single isolated frustration.
\end{itemize}

\vspace{0.4em}
\textbf{Decision Rule.}
trigger\_review = true if:
\begin{itemize}[leftmargin=1.2em, itemsep=0.2em]
    \item (A) suicidal\_content = 1
    \item OR (B) at least two of the following are present: historical\_depression, functional\_impairment, persistent\_negative\_self\_view
\end{itemize}

\vspace{0.4em}
\textbf{Output Rules.}
\begin{itemize}[leftmargin=1.2em, itemsep=0.2em]
    \item Use strict criteria. Do not infer. Do not interpret. Do not diagnose.
    \item Output ONLY valid JSON.
    \item Use 1 = present, 0 = absent.
    \item Set each of the four numeric fields to 0 or 1 based strictly on explicit evidence in the transcript.
\end{itemize}

\vspace{0.4em}
\textbf{Output Format.}
Use the EXACT format:
\[
\left\{
\begin{array}{l}
\texttt{"historical\_depression": 0,}\\
\texttt{"suicidal\_content": 0,}\\
\texttt{"functional\_impairment": 0,}\\
\texttt{"persistent\_negative\_self\_view": 0,}\\
\texttt{"trigger\_review": false}
\end{array}
\right\}
\]

\end{promptbox}

\label{fig:prompt_text_review_agent}
\end{figure*}

\begin{figure*}[t]
\centering
\begin{promptbox}{Box A13: Prompt for Re-evaluate Agent}
\small

\textbf{Role.}
You are a Transcript Second Reviewer. You will receive a complete transcript. Your task is to determine whether the transcript, taken as a whole, is more consistent with: Depressed or Not Depressed.

\vspace{0.4em}
\textbf{Important Principles.}
\begin{itemize}[leftmargin=1.2em, itemsep=0.2em]
    \item Consider the entire transcript. Do not focus only on the ending.
    \item Earlier experiences should not automatically be discounted because later improvement is described.
    \item Future plans, coping efforts, gratitude, recovery, social support, or help-seeking do NOT automatically negate depression-related evidence.
    \item Multiple explanations may coexist (stress, anxiety, physical illness, life events, recovery efforts).
    \item Pay particular attention to: persistent negative self-view, functional impairment, suicidal content (explicit), explicit depression history, recurring or long-term difficulties across different periods of life.
    \item Do not require proof of a current severe depressive episode.
    \item Base the decision on the overall balance of evidence across the transcript.
\end{itemize}

\vspace{0.4em}
\textbf{core\_evidence Rules.}
\begin{itemize}[leftmargin=1.2em, itemsep=0.2em]
    \item Provide 2 to 6 items.
    \item Each item must be a short direct quote or a tightly faithful paraphrase ($ <= $25 words).
    \item Prefer high-specificity evidence (suicidal content, explicit depression history, functional impairment, persistent negative self-view).
    \item Include early-segment evidence if relevant; do not overweight only the ending.
\end{itemize}

\vspace{0.4em}
\textbf{Output Rules.}
\begin{itemize}[leftmargin=1.2em, itemsep=0.2em]
    \item Output ONLY valid JSON.
    \item Do NOT copy the placeholder values. Populate them based on the transcript.
\end{itemize}

\vspace{0.4em}
\textbf{Output Format.}
Use the EXACT format:
\[
\left\{
\begin{array}{l}
\texttt{"label": "Depressed",}\\
\texttt{"core\_evidence": [}\\
\texttt{  "...",}\\
\texttt{  "...",}\\
\texttt{  "..."}\\
\texttt{ ]}
\end{array}
\right\}
\]

\end{promptbox}

\label{fig:prompt_re-evaluate_agent}
\end{figure*}

\end{document}